%% file: sn-article_v2.tex
\documentclass[referee,pdflatex,sn-mathphys-num]{sn-jnl}
\usepackage{graphicx}% Include figure files
\usepackage{multirow}%
\usepackage{amsmath,amssymb,amsfonts}%
\usepackage{amsthm}%
\usepackage{mathrsfs}%
\usepackage[title]{appendix}%
\usepackage{xcolor}%
\usepackage{textcomp}%
\usepackage{manyfoot}%
\usepackage{booktabs}%
\usepackage{algorithm}%
\usepackage{algorithmicx}%
\usepackage{algpseudocode}%
\usepackage{listings}%
\usepackage{dcolumn}% Align table columns on decimal point
\usepackage{bm}% bold math
\usepackage[acronym]{glossaries}
\usepackage{xr-hyper}
\usepackage{cleveref}
\usepackage[addedmarkup=colored]{changes}
\usepackage{soul}

\theoremstyle{thmstyleone}%
\theoremstyle{thmstyletwo}%

\theoremstyle{thmstylethree}%

\newacronym{EDX}{EDX}{energy dispersive X-rays}
\newacronym{MLE}{MLE}{maximum likelihood estimation}
\newacronym{STEM}{STEM}{scanning transmission electron microscopy}
\newacronym{EELS}{EELS}{electron energy loss spectroscopy}
\newacronym{FOV}{FOV}{field of view}
\newacronym{SBR}{SBR}{signal-to-background ratio}

\begin{document}

\title[Three-dimensional dopant distribution from a single, out-of-focus spectroscopic image]{Three-dimensional dopant distribution from a single, out-of-focus spectroscopic image}

\author[1]{\fnm{Michael} \sur{Deimetry}}\email{Michael.Deimetry@monash.edu}

\author[2,1]{\fnm{Timothy} \sur{Petersen}}\email{Timothy.Petersen@monash.edu}

\author[2,3]{\fnm{Matthew}  \sur{Weyland}}\email{Matthew.Weyland@monash.edu}

\author*[1]{\fnm{Scott} \sur{Findlay}}\email{Scott.Findlay@monash.edu}

\affil[1]{\orgdiv{School of Physics and Astronomy}, \orgname{Monash University}, \orgaddress{\country{Australia}}}

\affil[2]{\orgdiv{Monash Centre for Electron Microscopy}, \orgname{Monash University}, \orgaddress{\country{Australia}}}

\affil[3]{\orgdiv{Department of Materials Science \& Engineering}, \orgname{Monash University}, \orgaddress{\country{Australia}}}


\abstract{
Increasing miniaturization of nanoscale devices makes determining the three-dimensional distribution of dopants increasingly important. In out-of-focus spectroscopic scanning transmission electron microscopy images, we show that dopant atoms appear as disks with radii relating to dopant depth, allowing the three-dimensional distribution of dopants to be determined from simple geometric analysis of a single out-of-focus image. Through numerical experiments on Si-doped graphene we validate statistical detection and Bayesian inference methods to quantify confidence and uncertainties while making the analysis more robust at lower dose. We present proof-of-concept experimental data from an Fe-doped spinel, retrieving dopant depth to within around 2 nanometres.
}

% \keywords{keyword1, Keyword2, Keyword3, Keyword4}



\maketitle

\section*{Introduction}\label{sec:introduction}

As device scales shrink to nanometer dimensions, small numbers of dopant atoms play increasingly central roles – beneficial or otherwise – in areas including semiconductor technology \citep{koenraad2011single}, photovoltaics \citep{park2018point}, quantum emitters \citep{diler2020coherent} and quantum computing \citep{yang2023controlling}. Measuring the distribution of dopant atoms requires high spatial resolution in three dimensions and high sensitivity \citep{hunnestad2022atomic}. Scanning transmission electron microscopy (STEM), with angstrom lateral resolution and few nanometer depth resolution, offers routes to achieve this \citep{xu2020single}. 
Depth sectioning using annular dark field imaging has proven successful for localising heavy dopants in a light host crystal \citep{wang2004dopants,oh2008point,ishikawa2020three}. Recent developments in multislice ptychography provide an alternative approach to dopant imaging albeit via considerable post-processing \citep{chen2021electron,dong2025sub}.

Electron energy loss spectroscopy (EELS) can probe a wider range of dopant elements, confirm dopant species, and provide an indicator of dopant presence during acquisition. Single atom STEM-EELS imaging has been well demonstrated \citep{suenaga2000element,suenaga2009visualizing,ramasse2013probing,lin2015exploring}, %but does not provide depth information and requires very high dose. To address these challenges, we propose probing the 3D distribution of dopants via a \emph{single} defocused STEM EELS image. 
but does not provide depth information. Here we show that a single out-of-focus STEM EELS image does provide depth information, and can thus be used to characterize the 3D distribution of dopants. Depth-from-focus is a well-established family of techniques in light optics \citep{zheng2025depth}, but is routinely used in STEM only in so much as bringing a sample into focus involves intuitive and qualitative use of the underlying principle.

In this study, we introduce the depth from focus principle on ideal simulated data. Low ionization cross-sections have traditionally restricted single atom STEM EELS to very high dose imaging conditions. Through numerical experiments that incorporate shot noise and spectral background, we demonstrate that using statistical detection rather than imaging mitigates dose requirements, thereby increasing the diversity of samples that can be analysed. By applying the method to experimental data from an Fe-doped spinel, both lateral and depth resolution of a dopant were obtained with comparable precision to that predicted from simulation.

\section*{Results and discussion}\label{sec:results-discussion}

\subsection*{Geometric analysis for depth from focus}
% \subsection*{Mechanism of Depth Contrast Formation}

\begin{figure*}
    \centering
    \includegraphics[width=1\linewidth]{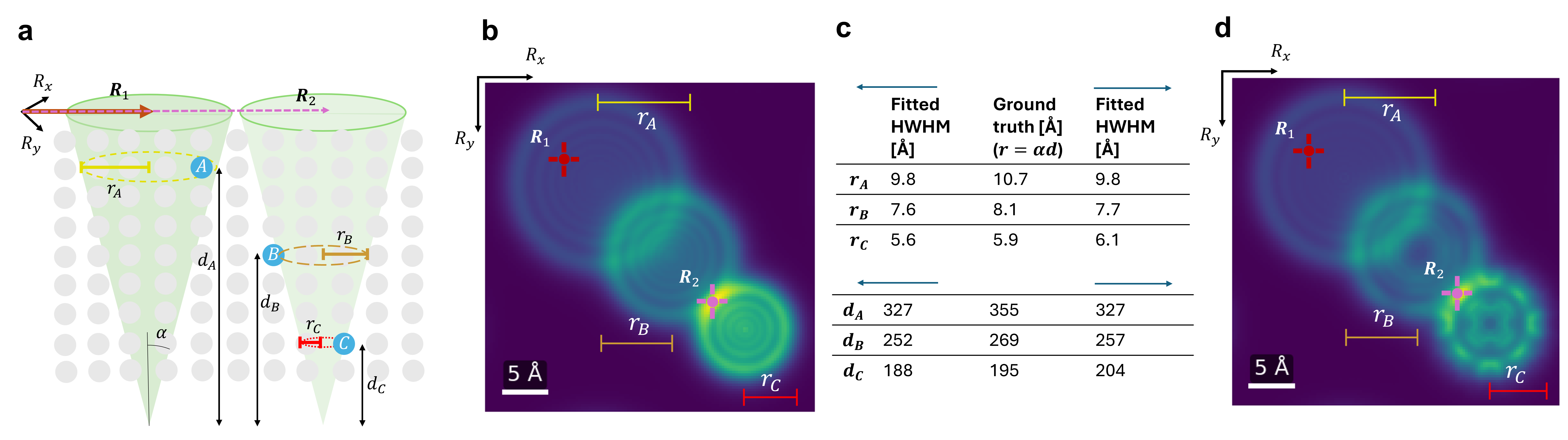}
    \caption{{\bf a} Schematic showing two scan positions of a highly defocused STEM probe on a sample containing three dopants labelled A, B and C. {\bf b} Simulated STEM-EELS image for the L$_{2,3}$ edge of three Fe dopants at different depths for 300keV electrons and $\alpha=30$mrad, excluding scattering from the host crystal. {\bf c} Table comparing disk radii fit using \cref{eq:sup_gaussian} and the dopant depths deduced via \cref{eq:rad} from the images in {\bf b} and {\bf d} against the ground truth values. {\bf d} As per {\bf b} but including scattering from the SrTiO$_3$ host crystal. Though {\bf a} is not to scale, the probe-position cross-hairs and disk radii labels in {\bf b} and {\bf d} correspond well enough with their counterparts in {\bf a} to elucidate the contrast formation mechanism.
    }
    
    \label{fig:showcase}
\end{figure*}

\Cref{fig:showcase}{\bf a} depicts a STEM scan for a highly defocused probe with focal point below the sample, across a structure containing three dopants (labelled A, B and C) at different lateral positions and depths. To elucidate the principle, we initially ignore the host crystal atoms and consider just the dopants. \Cref{fig:showcase}{\bf b} shows a simulated STEM-EELS image for the L$_{2,3}$ edge for a configuration of three Fe dopants with lateral separation around 10{\AA} and depth separation around 80{\AA}, with 300keV electrons and probe convergence semi-angle $\alpha=30$mrad. In atomic-resolution STEM EELS, atomic columns appear as narrow peaks when imaged in-focus, but the large defocus here blurs those peaks into diffuse disks with some ring-shaped internal structure. Nevertheless, the centre of each disk corresponds to the lateral location of the atom. Moreover, the disk radius (and to a lesser extent disk intensity) relates to the dopant depth.

This can be understood from simple geometric optics. Though not to scale, \cref{fig:showcase}{\bf a} conveys the main concepts. The geometric illumination cones for two probe positions, ${\bf R}_1$ and ${\bf R}_2$, are shown. If appreciable ionization only occurs from dopants located within the illumination cone, the STEM-EELS image for each dopant would be a disk of radius equal to that of the illumination cone at the dopant depth. (Since the net electron flux is conserved at each depth in the illumination cone, the flux density is inversely proportional to the disk area, explaining why in \cref{fig:showcase}{\bf b} larger disks have lower intensity.) In \cref{fig:showcase}{\bf b}, probe position ${\bf R}_1$ lies within the disk of radius $r_a$; in \cref{fig:showcase}{\bf a}, the illumination cone at probe position ${\bf R}_1$ encompasses only dopant A, with illumination cone radius at that depth being $r_A$. In \cref{fig:showcase}{\bf b}, probe position ${\bf R}_2$ lies in an overlap region between the disks with radii $r_B$ and $r_C$; in \cref{fig:showcase}{\bf a}, the illumination cone at probe position ${\bf R}_2$ encompasses both dopants B and C, with illumination cone radius $r_B$ and $r_C$ at their respective depths.

It follows immediately from the illumination cone geometry in \cref{fig:showcase}{\bf a} that the disk radius $r$ at some depth is related to the distance $d$ of that depth from the focal plane via 
\begin{equation}
r=\tan(\alpha)d\approx \alpha d,
    \label{eq:rad}
\end{equation}
with probe convergence semi-angle $\alpha$ typically being around 20-30mrad for atomic-resolution STEM justifying the small angle approximation. \Cref{eq:rad} converts disk radii measurements to relative dopant depths: depth-from-focus achieved from a single out-of-focus STEM-EELS image.

We demonstrate this on \cref{fig:showcase}{\bf b}, objectively determining disk radii using \gls{MLE} to fit the super-Gaussian model
\begin{equation}
    I({\bf R}) = \sum_i \frac{A_i}{A(r_i,n)} \exp\left[-\ln 2 \left(\frac{|{\bf R} - {\bf R}_i|}{r_i}\right)^{2n}\right],
    \label{eq:sup_gaussian}
\end{equation}
with ${\bf R}_i=(x_i,y_i)$ the centre of the $i^{\rm th}$ disk, $r_i$ its radius at half maximum, $n$ is the super-Gaussian order, controlling disk edge sharpness and $A(r_i,n)$ is the super-Gaussian area which ensures $A_i$ is the integrated electron counts. The sum is over all dopants in the image. 
This fitting strategy struggles if many disks overlap, and so is most suited to relatively dilute dopant distributions. The left and middle columns in \cref{fig:showcase}{\bf c} compare disk radii inferred from the MLE fit parameters of \cref{eq:sup_gaussian} to the STEM-EELS image in fig. 1{\bf b}, and the dopant depths then deduced via \cref{eq:rad} against the input (ground truth) values. For this idealised case, the deduced depths are accurate to within 2 to 3 nanometres.

Because the geometric model is somewhat simplistic, consider the 3D linear model \cite{xin2009aberration}, a 3D generalisation of the object function approximation \cite{cherns1973characteristic,rafferty1999towards} whereby the EELS image $ I({\bf R})$  is described by an overlap integral between the intensity of the probe wavefield $\psi_{\mathrm{probe}}$ and the summed intensity of the contributing transition potentials $H_{m0}({\bf R},z)$ \cite{dwyer_multislice_2005}
\begin{align}
I({\bf R}) &=  \iint |\psi_{\mathrm{probe}}({\bf R} - {\bf r},z)|^2 %\nonumber \\ & \qquad\qquad \times
\sum_{i,m} |H_{m0}({\bf r}-{\bf R}_i,z-z_i)|^2 dz \; d^2{\bf r} \nonumber \\ 
&\approx \sum_i \left[ \left|\psi_{\rm probe}({\bf R},z_i)\right|^2 \otimes_{\bf R} \sum_m \left|H_{m0}({\bf R}-{\bf R}_i,z_i)\right|^2 \right],\; 
\label{eq:conv_model}
\end{align}
where $\otimes_{\bf R}$ denotes convolution with respect to probe position and the second line assumes the inelastic transitions are localized to the dopant depths. As shown in supplementary \cref{sec:transition-potential-width}, the probe intensity at any given depth may have appreciable structure, as may the transitions potentials, but the convolution has less structure, particularly for the more delocalised interactions at lower ionization energies. Residual structure from the convolution explains the ring-like structure within the disks in \cref{fig:showcase}{\bf b}. We have empirically found \cref{eq:sup_gaussian} to be a good compromise between model simplicity and accurate fitting to \cref{eq:conv_model}, with the radius at half maximum corresponding to the dopant depth via \cref{eq:rad} (see supplementary \cref{sec:fwhm-justification}). One could fit a model more closely matching \cref{eq:conv_model}, but, as real data are more likely limited by noise than model mismatch, the simpler form of \cref{eq:sup_gaussian} should suffice. 
Moreover, the empirical \cref{eq:sup_gaussian} does not presume knowledge of the dopant species, EELS cross section, or detailed probe distribution.

%\Cref{fig:showcase}{\bf b} is an idealisation, assuming infinite dose and ignoring the host crystal. To move beyond the idealisation of \cref{fig:showcase}{\bf b}, let us explore individually the impact scattering through the host crystal and finite dose have upon this method. 
In \cref{fig:showcase}{\bf b} we ignored the host crystal. To explore the impact scattering through the host crystal has upon this method, \cref{fig:showcase}{\bf d} shows the simulated STEM-EELS image for the same Fe dopants within a 195Å-thick SrTiO$_3$ sample. Dopant $C$, deepest in the crystal where the probe wavefield has been notably deformed by multiple electron scattering, shows the largest change in contrast. Conversely, dopant $A$, nearest the entrance surface where the probe wavefield has been little deformed by scattering, shows little change in contrast: multiple scattering of the post-ionization inelastic wavefield redistributes intensity across the detector but, with a large collection angle, does not much change the measured intensity. Thus despite the presence of the host crystal, the \gls{MLE} parameters of \cref{eq:sup_gaussian} (right column in \cref{fig:showcase}{\bf c}) remain within a few nanometers of the true depths.

% ======================================================
% ================== inference methods ===================
% ======================================================
\subsection*{Statistical inference on numerical experiments}
\begin{figure}
    \centering
    \includegraphics[width=0.7\linewidth]{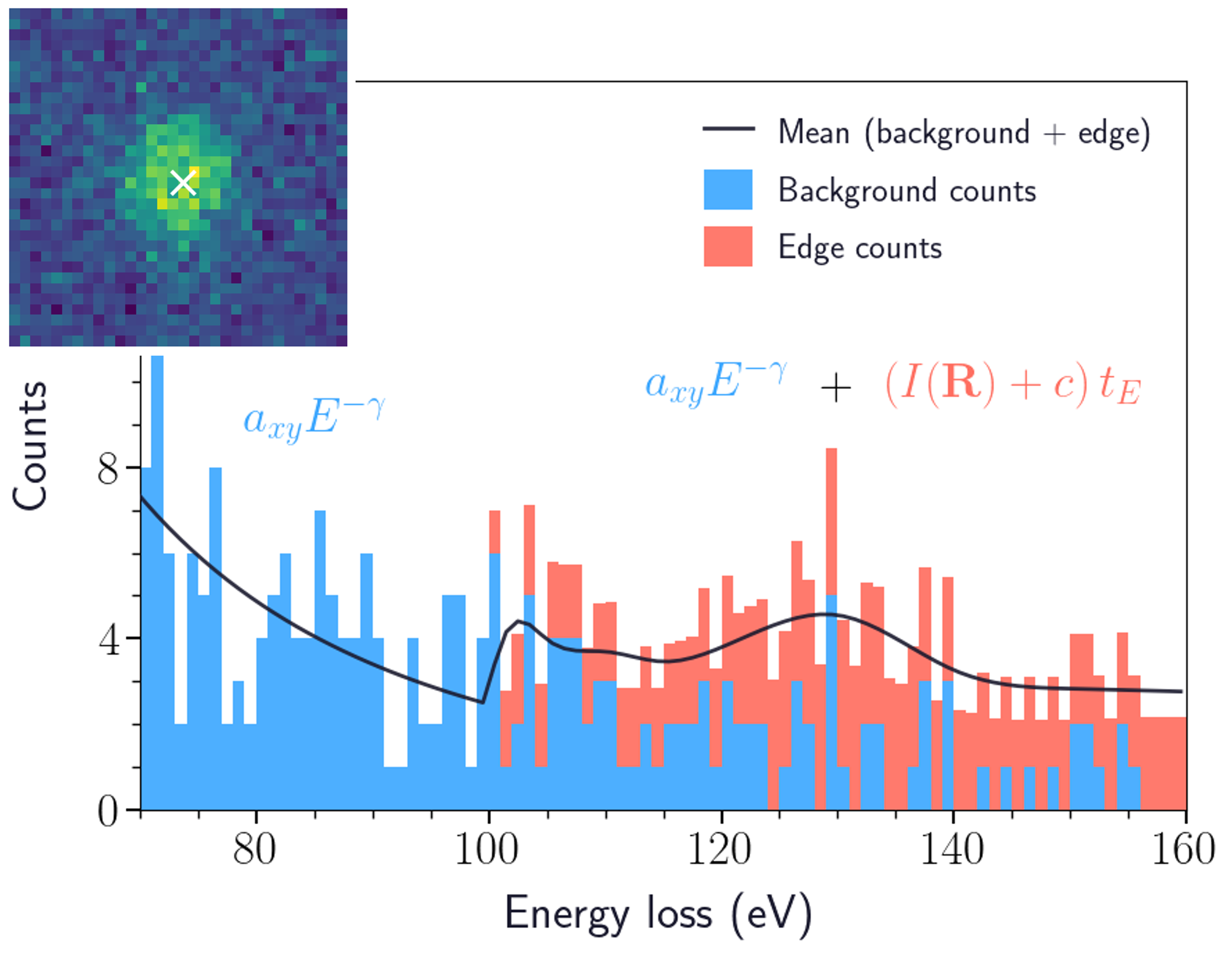}
    \caption{Simulated STEM-EELS image, after background subtraction, for the L$_{2,3}$ edge of a single Si atom, assuming 80keV electrons and $\alpha=30$mrad, including Poisson noise appropriate to an expected 5,000 counts across the image. The spectrum from a single pixel of that image shows the background and edge shapes used and how shot noise affects both, assuming an edge signal to background ratio of 1. Colour is used for illustrative purposes only; background and edge counts can only approximately be deduced in experiment via background fitting and subtraction.%counts=5k,SBR=1, inset shows just core-loss counts, background is included. dose=2e6
    }
    \label{fig:ex_spectrum}
\end{figure}

To convey the depth-from-focus principle, \cref{fig:showcase}{\bf b} and {\bf d} assume infinite dose and show only intensity resulting from ionization events, making the disks clearly evident. However, experimental data will look more like the image in \cref{fig:ex_spectrum}: finite dose, low ionization cross-section and spectral background make for noisy data in which it may be hard to discern any disks present. \Cref{fig:ex_spectrum} also shows an example energy loss spectrum with shape modelled from the literature \cite{ramasse2013probing} and counts given by \cite{verbeeck2004model}
\begin{equation}
    D_{xyE} \sim \text{Poisson}\big(I_{xyE}=\underbrace{(I({\bf R})+c)\,t_E}_{\text{core-loss signal}} + \underbrace{a_{xy}b_E}_{\text{background}} \big)  \;,
    \label{eq:eels_model}
\end{equation}
indexed by probe-position coordinates $(x,y)$ and energy-loss $E$, and where $I({\bf R})$ is the total core-loss counts at pixel ${\bf R}=(x,y)$, $t_E$ is the normalized edge shape, $a_{xy}$ is the amplitude of the background with power-law shape $b_E=E^{-\gamma}$, and $c$ is a constant signal offset. Note that the edge and background shapes do not vary across pixels. Poisson random deviates model fluctuations away from the expectation value $I_{xyE}$ due to shot noise. %While the spectrum in \cref{fig:ex_spectrum} distinguishes background and edge counts by colour for conceptual clarity, that is not possible experimentally for individual electrons.

Our case study for exploring the impact of noise is the L$_{2,3}$ edge from a single Si dopant in graphene. This atomically thin sample eliminates multiple electron scattering effects. We assume a 30mrad convergence semi-angle and 20nm defocus, producing a disk radius of 6{\AA}, and a 1{\AA} probe step size. The background amplitude at each probe position is independently drawn from a Gaussian distribution with standard deviation 20\% of the mean value, a more challenging scenario than likely applies for a highly defocused probe even given the large probe step size. %We model the per-pixel background amplitude variation by $a_{xy} \sim \mathcal{N}(\mu=a, \sigma=0.2 a)$ for some reference amplitude, $a$. This implies background variation across the FOV is not smooth, as would be expected in practice, which constitutes a harder case study.
 %We expand the applicability of this case study in two ways. First, we vary the total core-loss signal counts (across all probe positions for a single dopant). Since the analyses presented work primarily on EELS data cubes we would expect similar behaviour for other elements and edges based on the total number of counts, though how that translates to dose depends on the ionization cross-section and probe sampling considered. Second, we also vary the \gls{SBR} --- the background depends on the host crystal and increases with sample thickness. 
To make our findings applicable beyond this specific case study, we independently vary the total core-loss signal counts in the image and the \gls{SBR}. The former allows the reader to estimate the dose that might be required for other elements and edges based on their ionization cross-section. The latter allows for the background varying with host crystal and increasing with sample thickness. We explore three analysis methods, with the goal of quantifying statistical confidence and increasing robustness to noise and so making analysis possible at lower doses.
The first two methods primarily detect dopants, while the third further yields quantitative depth estimates.
%\hl{Only the third method, Bayesian inference, will yield quantitative depth estimates.}

The first method is conventional background subtraction. Using a pre-edge window, we first estimate $\hat{\gamma}$ via least-squares on the spectrum summed over the field of view (adopting the convention of denoting \gls{MLE} results with hats). As derived in supplementary \cref{sec:background-amplitude-mle}, \gls{MLE} for $a_{xy}$ gives
\begin{align}
\hat{a}_{xy}
= \frac{1 }{B} \sum_{E \in \mathrm{bkg}} D_{xyE} , \quad \text{Var}(\hat a_{xy}) \approx \frac{\hat a_{xy}}{ B }
       \label{eq:amp_mle}
\end{align}
where $B= \sum_{E \in \mathrm{bkg}} \hat b_E$ and all sums are over the pre-edge window. Observe the variance is inversely proportional to the total pre-edge window background counts, and so can be reduced by increasing the pre-edge window energy width. %Hence, for a pre-edge window of 100eV the variance would be negligible.
The distribution of $\hat{a}_{xy}$ estimates with pre-edge window length is explored in supplementary \cref{sec:background-amplitude-mle}.

\Cref{fig:inference_showcase} shows simulations for the Si-L$_{2,3}$ case study from two different signal levels (1,000 counts, here implying a dose of $4 \times 10^5 \; {\rm e}^-/{\rm \AA}^2$, and 100 counts, here implying a dose of $4 \times 10^4 \; {\rm e}^-/{\rm \AA}^2$) and three different SBR values (1, 0.25 and 0.0625). The first row shows the results of background subtraction. Only in \cref{fig:inference_showcase}{\bf a}, for 1,000 counts and SBR=1, is the disk perceptible by eye. Denoising by Gaussian blurring makes disk-like features more evident in some of the other figures (see supplementary \cref{sec:gaussian-blurring}), but we must introduce some metric of significance to inform whether such processing is trustworthy.

This second method is a frequentist score test consisting of a hypothesis test (the null hypothesis being no signal) on a per-pixel basis that template matches the background-subtracted residual with the expected edge shape. As derived in supplementary \cref{sec:score-test}, the resulting standardized score map is
\begin{equation}\label{eq:zmap}
    z_{xy}
    = \frac{\sum_{E \in \text{edge}}
            (D_{xyE} - \hat a_{xy}\hat{b}_E)\,\dfrac{\hat t_E}{\hat{b}_E}}
           {\sqrt{\hat a_{xy}
            \sum_{E \in \text{edge}}
            \dfrac{\hat t_E^2}{\hat{b}_E}}}.
\end{equation}
A large positive value indicates the residual spectrum at that pixel resembles the edge shape more than would be expected from noise alone.  Applying a modest Gaussian blur to $z$-score maps is again useful to improve the visibility of disk-like features.

Smoothed $z$-score maps are shown in the second row of \cref{fig:inference_showcase}. In \cref{fig:inference_showcase}{\bf g} (1,000 counts and SBR=1) the disk is now clearly visible at $z_{xy}\gtrsim 2$ (i.e. around 2$\sigma$ significance). Moreover, we now have good confidence of dopant presence in \cref{fig:inference_showcase}{\bf h} (1,000 counts and SBR=0.25) and \cref{fig:inference_showcase}{\bf j} (100 counts and SBR=1), though the disk radius is increasingly hard to discern. \Cref{fig:inference_showcase}{\bf i} and {\bf l} have maxima suggestively near the dopant location, but their lower $z$-score reduces detection confidence. This analysis is fast enough to be useful for guiding an experimental session. The more reliable 1,000 count case is lower dose than most single atom imaging \cite{suenaga2000element,suenaga2009visualizing} and comparable to ptychography \cite{chen2021electron}. To explore whether the dose required can be reduced still further, broadening the applicability of the method, we turn to a model-based inference strategy that considers the STEM-EELS data cube holistically.

This third method is Bayesian inference for the model parameters in \cref{eq:sup_gaussian} and \cref{eq:eels_model}:
\begin{align}
    \Theta=\{\underbrace{ \{A_i,r_i,x_i, y_i\},c }_{\text{disk params.}}, \underbrace{a_{xy},\gamma}_{\text{background params.}}\}
\end{align}
(the super-Gaussian order $n$ can be inferred or fixed, depending on whether the disk sharpness is estimated in advance). 
% As detailed in supplementary material, t
The method involves estimating the posterior $\mathcal{P}(\Theta \mid D, M)$ (with $D$ denoting
the observed data and $M$ encoding prior knowledge about model parameters $\Theta$) via Bayes’ theorem
\begin{align}
\mathcal{P}(\Theta \mid D, M) = \frac{\mathcal{L}(D \mid \Theta, M)\,\pi(\Theta \mid M)}{\mathcal{Z}(D \mid M)} \;,
\label{eq:bayes_thm}
\end{align}
where $\pi(\Theta \mid M)$ denotes the prior distribution of the model parameters $\Theta$, the Poisson likelihood for the EELS data cube 
\begin{align}
\mathcal{L}\!\left(
D
\mid
\Theta,
M
\right) 
=
\prod_{\substack{x,y\\ E\in\mathrm{edge}}}
\frac{
I_{xyE}^{D_{xyE}} e^{-I_{xyE}}
}{
D_{xyE}!
} \;,
\label{eq:ll_img}
\end{align}
gives the probability of the observed data given the model parameters and accounting for Poisson shot noise, and $\mathcal{Z}(D \mid M)$, known as the evidence, is a normalization constant obtained by marginalizing over the model parameter space. Since inference with many parameters can be slow, we infer background parameters $a_{xy}$ and $\gamma$ through \gls{MLE} on the pre-edge window, reducing the problem dimensionality to that of the disk parameters. The edge shape can be obtained in the same manner as in the frequentist analysis. The super-Gaussian order $n$ can be fixed based on simulated STEM-EELS image references or inferred from the data, though note that a poor estimate may lead to model mismatch and incorrect depth estimates.

Marginalised posteriors for the joint disk centre coordinates and dopant depth (deduced from disk radius via \cref{eq:rad}) are shown in the third and fourth rows, respectively, of \cref{fig:inference_showcase}, with super-Gaussian order inferred from data. Fuller visualisation of the posterior for other parameters are given in supplementary \cref{sec:verifying-inference}. At 1,000 counts ($4\times 10^5 \; e^-/\mathrm \AA$) the peaks in the posteriors for disk centre and radius agree fairly well with the true values for all SBR considered, though lower SBR broadens the posterior, increasing the uncertainty. For SBR=1 the dopant depth posterior is offset from the ground truth due to mismatch between the model of \cref{eq:sup_gaussian} and pronounced disk tail due to the delocalized Si-L$_{2,3}$ interaction (though this could be overcome using a model closer to \cref{eq:conv_model}). At lower dose and SBR, model mismatch is masked by the uncertainty from noise.

For the lower dose of 100 counts ($4\times 10^4 \; e^-/\mathrm \AA$),  reducing SBR has a far greater influence on the posteriors. 
At SBR=1, the posterior distribution variance about the peak is small, allowing reasonably accurate dopant depth estimation. At SBR=0.25, a peak is still discernible but is considerably wider, suggesting only a partial update over the prior. At SBR=0.0625, the posterior closely mirrors the prior distribution, implying that the data do not meaningfully constrain the disk parameters beyond prior assumptions: dopant location and depth determination fails when background noise swamps a weak edge signal.

\begin{figure*}
    \centering
    \includegraphics[width=1\linewidth]{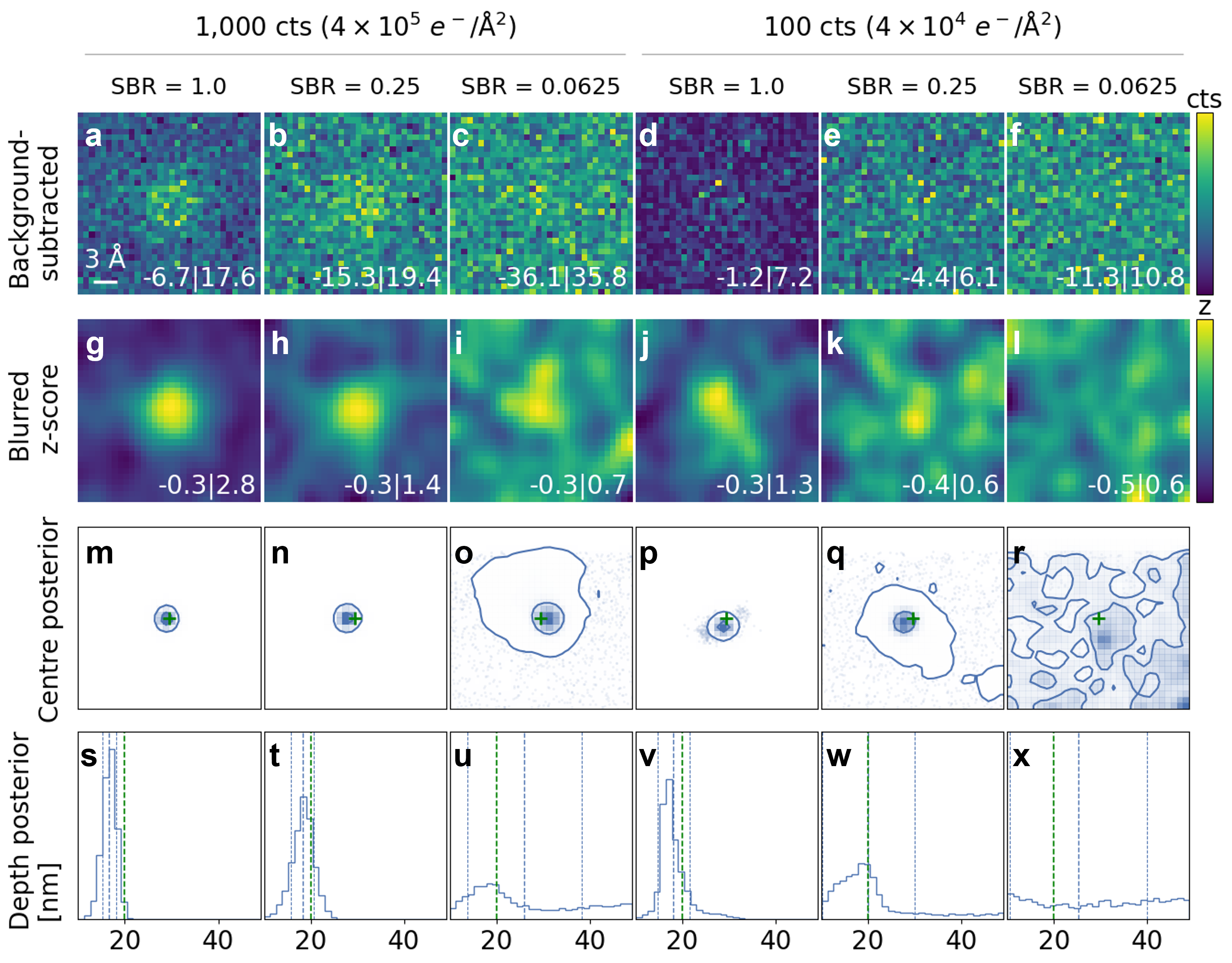}
    \caption{Comparison of three analysis approaches --- background subtraction (first row), standardized score map of \cref{eq:zmap} (second row) with 2${\rm\AA}$ standard deviation Gaussian blur,  lateral atom position (third row) via Bayesian inference using \cref{eq:bayes_thm} and marginalised posteriors for dopant depth (fourth row) --- on simulated STEM-EELS image for the L$_{2,3}$ edge of a single Si atom in graphene for difference combinations of total signal counts and SBR. 
    Rows 1 and 2 use a common colour scale, with the corresponding minimum and maximum values shown in the bottom-right corner of each subplot as \texttt{\textless min\textgreater} $\mid$ \texttt{\textless max\textgreater}.
    The analyses assume a pre-edge window from 50-90 eV and a signal window from 100-120 eV. True values are indicated by the green crosses in the plots of row 3, and green dashed lines in the plots of row 4. 
    %stopping condition: dlogz < 0.05,
    %signal window=[100, 120]
    %fov window = [50, 160]
    %pre-edge window = [50, 90]
    }
    \label{fig:inference_showcase}
\end{figure*}

% signal_window=[100, 120]
% background_window = [50, 160]
% pre_edge_window = [50, 90]

% make consistent I and A
\subsection*{Experimental demonstration}
\begin{figure*}
    \centering
    \includegraphics[width=1\linewidth]{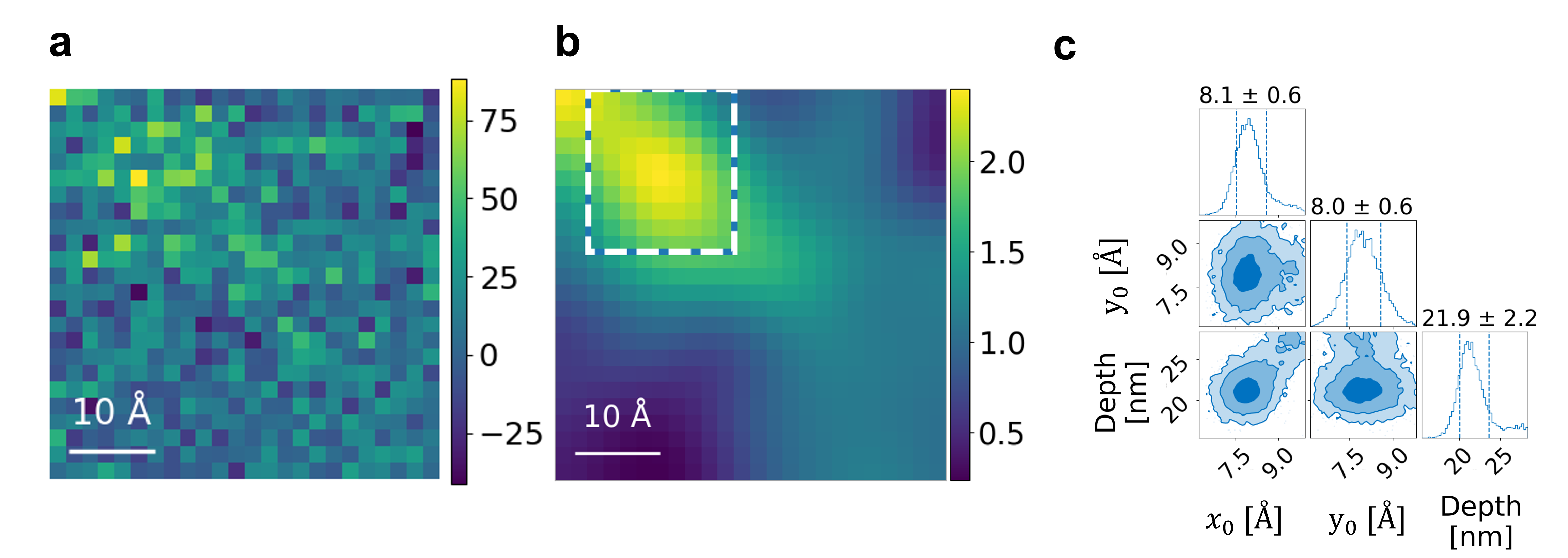}
    \caption{{\bf a} Background-subtracted experimental STEM-EELS image for the Fe L$_{2,3}$ edge in Fe-doped MgAl$_2$O$_4$, with 300keV electrons, $\alpha=30$mrad and a pre-edge window of 620-680eV. {\bf b} Corresponding standardized score map of \cref{eq:zmap} with 5${\rm\AA}$ standard deviation Gaussian blur. {\bf c} Marginalized posteriors from Bayesian inference on the region-of-interest highlighted in {\bf b}. 
    %background subtraction done over 620-680eV. Edge assumed in region 690-710eV. Discrepancy in edge onset is suspected to be due to the non-linear dispersion across the energy-loss spectrum from the large dispersion of 0.9eV per channel used. 45x45$\AA$ FOV. Reduced FOV is 19x17$\AA$. Posterior centre relative to reduced FOV top left corner. Dopant depth in nm, centre in A.
    }
    \label{fig:experiment_analysis}
\end{figure*}

As a proof-of-principle experimental demonstration of this approach, out-of-focus STEM-EELS data was acquired on an Fe-doped MgAl$_2$O$_4$ spinel sample viewed along [001] using 300keV electrons, a 30mrad probe convergence semi-angle and a step size of $2\mathrm{\AA}$ (see Methods for further details). The sample was underfocused by 20nm, corresponding to an expected increase in the dopant disk radius of approximately 6{\AA}. The dose of $6\times 10^4 ~e^-/\mathrm{\AA}^2$ chosen for adequate EELS signal led to appreciable beam-induced sample damage (a post-acquisition annular dark field survey is shown in supplementary \cref{sec:examining-fov}) and many Fe atoms appeared mobile under the beam. However, the experimental results in \cref{fig:experiment_analysis} show a case where one dopant was stable for most of the acquisition.

\Cref{fig:experiment_analysis}{\bf a} shows the background-subtracted Fe L$_{2,3}$ STEM-EELS image. Although the upper-left region appears brighter than the surrounding background, no distinct dopant disk is evident. However, the blurred $z$-score map in \cref{fig:experiment_analysis}{\bf b} clearly reveals a disk in this region. %A second disk is also visible near the upper-left corner of the field of view, which is consistent with HAADF imaging extending beyond EELS acquisition windows (see supplementary \cref{sec:examining-fov}). 
A second disk is also visible near the upper-left corner of the field of view, which is consistent with the survey annular dark field image taken over a wider field of view (see supplementary \cref{sec:examining-fov}). A faint diffuse signal is also present across much of the field of view, particularly in the right-hand region, where the $z$-score indicates significant departures from the null hypothesis. These are corroborated by spectra from probe-position binning of the STEM-EELS data, which show Fe peaks consistent with the $z$-score map (see supplementary \cref{sec:examining-fov}). We attribute these streaks to a dopant atom becoming dislodged during acquisition.

To avoid fitting streak-like features that are not well-captured by the forward model, Bayesian inference was performed on the cropped region outlined in blue and white containing the most prominent dopant. The super-Gaussian order was fixed from simulated reference. The posterior over the disk parameters is shown in \cref{fig:experiment_analysis}{\bf c}. The disk is estimated to be centred within the cropped field of view and has a radius corresponding to a dopant located approximately 22nm from the in-focus plane. Using position-averaged convergent beam electron diffraction pattern measurement, the sample thickness is estimated to be approximately 10nm (see supplementary \cref{sec:examining-fov}). Given the STEM-EELS image was acquired with 20nm underfocus relative to the annular dark field image, for which optimal contrast for this thin specimen occurs with a few nanometers underfocus of the entrance surface, the dopant is inferred to be close to the specimen entrance surface. This agreement was not imposed in the prior, as the disk radius prior was uniformly set over $0$-$8.5\mathrm{\AA}$.

The experimental SBR is around $0.1$ and the signal count is around $ 1000$ counts (see supplementary \cref{sec:verifying-inference}), making it broadly comparable to the simulations in \cref{fig:inference_showcase}{\bf n,o,t,u}. Consistent with this, the widths of the posteriors are indeed similar between these two cases, despite the difference in dopant species. In particular, the dopant depth uncertainty is on the order of 1nm. Further corroboration of the inferred parameters is given in supplementary \cref{sec:verifying-inference}.

In summary, simulations and proof-of-principle experimental results show that from a single out-of-focus STEM-EELS image it is possible to determine dopant distribution in three dimensions. Despite the low count rates for single atom spectroscopic imaging, statistical detection and Bayesian inference make the analysis possible at doses that are achievable for many materials science specimens. For the 300keV and 30mrad probes considered here, angstrom-scale lateral resolution and few nanometer depth resolution are  possible.  Improved resolution may be possible with higher convergence angle probes \cite{ishikawa2020three} or at higher signal-to-noise. Since defocused acquisition reduces the dose rate, this imaging strategy may suppress damage relative to in-focus single atom EELS imaging 
in materials where damage is dose-rate limited.

\section*{Methods}\label{sec:methods}

\subsection*{Sample preparation}

An Fe-doped MgAl$_2$O$_4$ crystal was crushed to create small shards using a zirconia mortar and pestle to avoid silicate contamination, with the sample dispersed in pure ethanol. The settling time was shortened to drop cast larger shards, exceeding several microns, onto a holey carbon film supported on a 400 mesh 3.05mm copper grid, to enable easier tilting onto crystal zone axes.

To diminish experimental artefacts, several additional preparation steps were undertaken. Firstly, the holey carbon film was pre-cleaned using pure ethyl acetate to remove hydrocarbons from residual polyvinyl formal (Formvar). To improve thermal and electrical conductivity, an AuPd sputter target was employed to half-coat the holey carbon support, prior to drop casting, giving rise to a continuous gradation in the concentration of AuPd nanocrystals, ranging from nascent to a continuous AuPd film, across the grid. Lastly, to ameliorate specimen instability, plasma cleaning of the drop-cast specimen was avoided. Instead, the sample was heated to 120°C ex-situ, effecting an ultrathin conductive coating of carbon and removal of moisture.

\subsection*{Microscope and acquisition details}

The sample was characterised using a Thermo Scientific Spectra  $\varphi$, a monochromated, dual aberration corrected field emission gun transmission electron microscope. The EEL spectra were acquired using a Gatan Continuum K3 HR spectrometer, fitted with a K3 direct electron detection camera. A spectrometer collection angle of 69mrad was used. Spectra were acquired in single EELS mode to reduce dose overhead. A dispersion of 0.9eV per pixel was used to maximise signal per bin, with an energy loss range from ~370eV to ~3100eV, collecting not only the Fe L$_{2,3}$ edge shown here, but also the O-K, Mg-K and Al-K edges. An exposure time of 20ms per spectra was used. Suitable areas for study were screened for thickness by collection of $t/\lambda$ maps. More precise thickness measurement was carried out by matching energy filtered position averaged convergent beam electron diffraction (PACBED) patterns  (shown in \cref{fig:fov_exam} of \cref{sec:examining-fov}), acquired using a Gatan K3 direction electron detection camera, to simulated patterns.

\subsection*{Image simulations}

Electron scattering simulations were performed using the multislice method as implemented in the open-source Python package py\_multislice~\cite{brown2020python,brown_py_multislice}, which supports core-loss simulations. A forked version of the repository \cite{md_py_multislice_fork} was used, extended with custom functionality for substituting dopant atoms into the simulation supercell. Simulations were run in a double channelling configuration, in which dynamical scattering of the beam electron is included both before and after the core-loss event. Absorptive potentials were used to account for the depletion of the elastic wave by thermal diffuse scattering.

\subsection*{Statistical inference}

All posterior distributions shown were generated using nested sampling via the Bilby library \cite{ashton2019bilby}, which provides an interface built on Dynesty's implementation \cite{sergey_koposov_2026_21417233,2020MNRAS.493.3132S}. The nested sampling stopping criterion was chosen when $\log \mathcal Z < 0.05$.

\section*{Data availability}\label{sec:data}
The simulated and experimental data for this paper can be accessed via the Monash University Bridges repository \cite{data_and_code}.

\section*{Code availability}\label{sec:code}
The analysis code for this paper can be accessed via the Monash University Bridges repository \cite{data_and_code}.

\bmhead{Acknowledgements}\leavevmode\\
The authors thank Dr Peter R Miller for supplying the Fe-doped spinel sample used for the defocused STEM-EELS experimental demonstration and Julian Ratcliffe for developing and preparing the half-coated holey carbon specimen supports, designed for aberration-corrected microscopy. This research was supported by the Commonwealth through an Australian Government Research Training Program Scholarship [DOI: https://doi.org/10.82133/C42F-K220]. This research was partially funded by the Australian Government through the Australian Research Council (project number DP250100935). The authors acknowledge the use of the instruments and scientific and technical assistance at the Monash Centre for Electron Microscopy, Monash University, a Microscopy Australia (ROR: 042mm0k03) facility supported by NCRIS. This research used equipment funded by Australian Research Council grant LE170100118.

\bmhead{Author contributions}\leavevmode\\
MD and SF designed the study and wrote the manuscript. MD developed the data analysis theory and code, and performed the analysis on both simulated and experimental data. MW and TP performed the STEM experiment. All authors discussed the analysis and results, and commented on the paper.

\bmhead{Competing interests}\leavevmode\\
The authors declare no competing interests.

\newpage

\backmatter

\bmhead{Supplementary information}

% If your article has accompanying supplementary file/s please state so here. 
% Please refer to Journal-level guidance for any specific requirements.

% Section 1
\include{supplementary/disk_structure_with_transition_potential_width}

% Section 2
\include{supplementary/FWHM_Justification}

% Section 3
\include{supplementary/Background_Amplitude_using_MLE}

% Section 4
\include{supplementary/Gaussian_Blurring_Effects_on_Inference}

% Section 5
\include{supplementary/score_test_derivation}

% Section 6
\include{supplementary/inference_strategy}

% Section 7
% \include{supplementary/Experimental Details}

% Section 8

\include{supplementary/Examining_FOV}

% Section 9
\include{supplementary/Verifying_Inference_Results}

\bibliography{sn-bibliography}% common bib file

\end{document}

% \preprint{APS/123-QED}

\title{Supplementary Material for ``Three-dimensional dopant distribution from a single, out-of-focus spectroscopic image''}% Force line breaks with \\

\author{Michael Deimetry}
 \affiliation{School of Physics and Astronomy, Monash University, Australia}

\author{Timothy C. Petersen}
 \affiliation{Monash Centre for Electron Microscopy, Monash University, Australia}

\author{Matthew Weyland}
 \affiliation{Monash Centre for Electron Microscopy, Monash University, Australia}
 \affiliation{Department of Materials Science \& Engineering, Monash University, Australia}

\author{Scott D. Findlay}
 \affiliation{School of Physics and Astronomy, Monash University, Australia}

\date{\today}

\maketitle

% Section 1
\include{supplementary/disk structure with transition potential width}

% Section 2
\include{supplementary/FWHM Justification}

% Section 3
\include{supplementary/Background Amplitude using MLE}

% Section 4
\include{supplementary/Gaussian Blurring Effects on Inference}

% Section 5
\include{supplementary/score-test derivation}

% Section 6
\include{supplementary/inference strategy}

% Section 7
\include{supplementary/Experimental Details}

% Section 8
\include{supplementary/Examining FOV}

% Section 9
\include{supplementary/Verifying Inference Results}

%% file: supplementary/disk_structure_with_transition_potential_width.tex
% \section{Disk Intensity vs Defocus and Core-Loss Edge}
\section{Disk uniformity in out-of-focus STEM-EELS images of dopants}\label{sec:transition-potential-width}

\begin{figure}
    \centering
    \includegraphics[width=.9\linewidth]{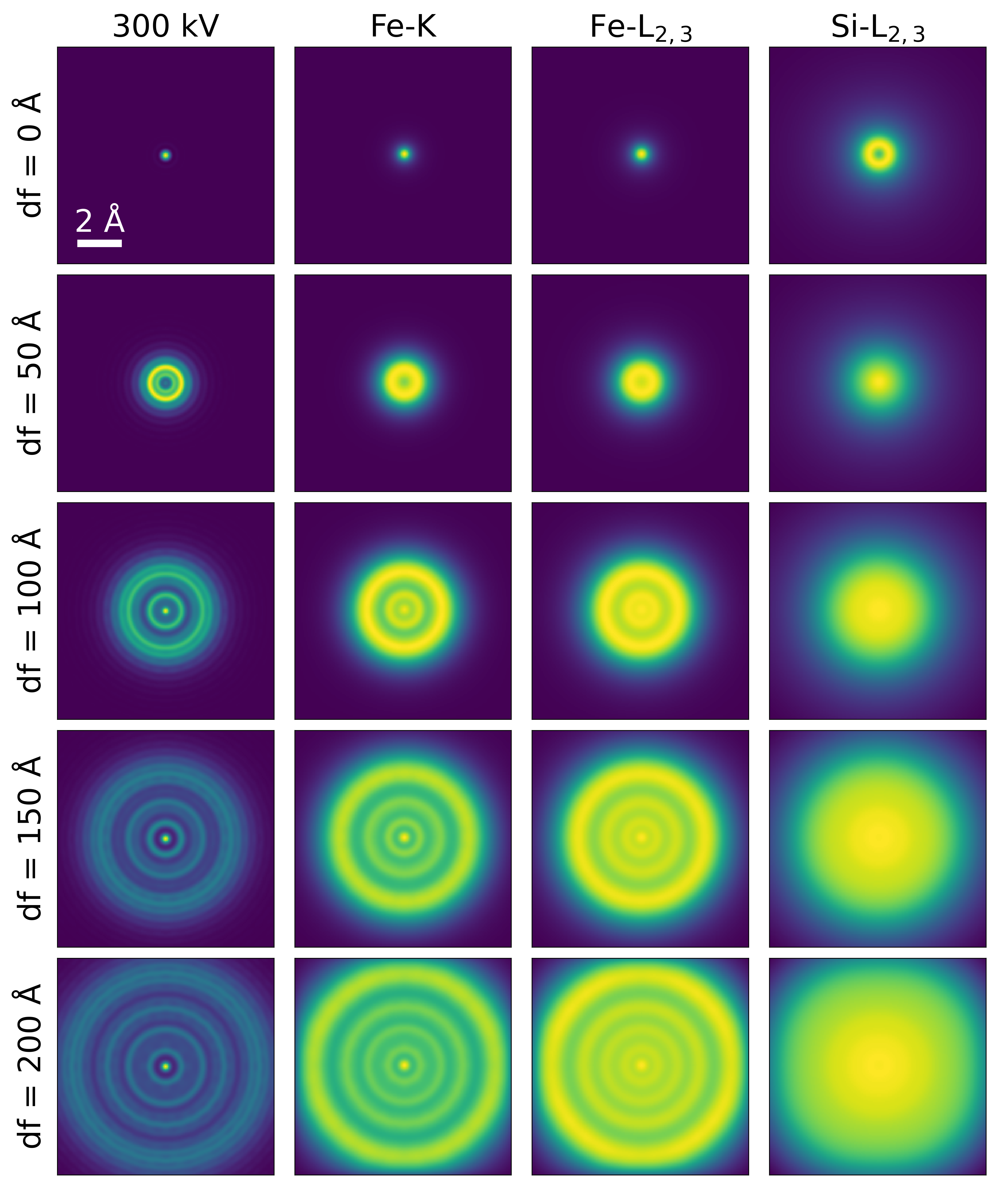}
    \caption{Simulated probe intensity and single-atom STEM-EELS images as a function of defocus (rows) for 300\,keV electrons with a 30\,mrad convergence semi-angle and 5\,eV energy window. (left column) Electron probe intensity, showing a localised in-focus peak that spreads and develops ring-shaped internal structure with increasing defocus. (remaining columns) Simulated STEM-EELS images for the Fe-K, Fe-L\textsubscript{2,3}, and Si-L\textsubscript{2,3} edges. The internal ring structure is progressively washed out from left to right as increasingly wide convolution kernels increasingly suppress fine structure, producing more uniform disks but with more blurred disk edges.}
    \label{fig:probe_vs_df_edge}
\end{figure}

The super-Gaussian model in \cref{eq:sup_gaussian} assumes that the intensity within a disk is perfectly uniform. While this would be true in the geometric model, it is not true in the wave-optical picture that is the more accurate description of electron wavefields in high-energy transmission electron microscopy. %That model applied to STEM results in non-uniform intensity distributions within a given defocus plane.
This is evident in the first column of \cref{fig:probe_vs_df_edge}, which shows simulated probe intensities as a function of defocus, assuming 300 keV electrons and a 30 mrad probe convergence semi-angle. The in-focus probe has a localised peak which then spreads and develops ring-shaped internal structure as defocus increases.

\begin{figure}
    \centering
    \includegraphics[width=0.5\linewidth]{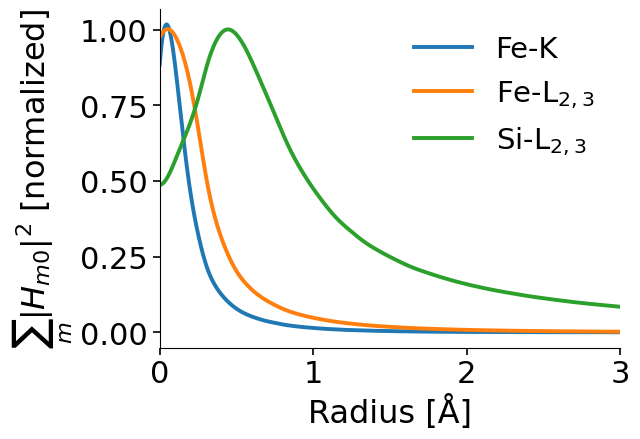}
    \caption{Normalized summed transition potential intensity profiles $\sum_m |H_{m0}|^2$ as a function of radius for the Fe-K, Fe-L\textsubscript{2,3}, and Si-L\textsubscript{2,3} edges. The profile narrows with increasing binding energy, with Fe-K being the most localised and Si-L\textsubscript{2,3} the least localised.}
    \label{fig:Hn0_profiles}
\end{figure}

The second through fourth columns show simulated single-atom STEM-EELS images for the Fe-K edge, Fe-L$_{2,3}$ edge and the Si-L$_{2,3}$, using the convolution model of \cref{eq:conv_model} of the main manuscript whereby the final EELS image is a convolution of the electron probe with a transition potential. The corresponding transition potential kernel profiles are shown in  \cref{fig:Hn0_profiles}, and as the binding energy decreases, from 7.1keV for Fe-K to 700eV for Fe-L$_{2,3}$ to 100eV for Si-L$_{2,3}$, the transition potential kernel widens. %The leftmost of these columns (Fe-K) has the narrowest kernel, while the rightmost (Si-L23) has the widest. 7112eV for Fe-K
This explains why the internal disk structure in the STEM-EELS images in the second through fourth columns of \cref{fig:probe_vs_df_edge} reduces: the wider the transition potential kernel, the more the internal structure of the probe cross-section is washed out by the convolution, making the disk interior more uniform. 

Note that visual disk uniformity does not necessarily improve the accuracy of the super-Gaussian model: the increasingly blurred edges themselves become a source of model mismatch that can limit depth resolution. In cases where model mismatch is expected to be appreciable, accuracy can be improved by instead fitting the more accurate convolution model, albeit this requires more elaborate simulations and increased computational cost.

% TODO: normalize all potentials?

%% file: supplementary/FWHM_Justification.tex
\section{Justification for using super-Gaussian HWHM as disk radius}\label{sec:fwhm-justification}

\begin{figure}[h!]
    \centering
    \includegraphics[width=0.5\linewidth]{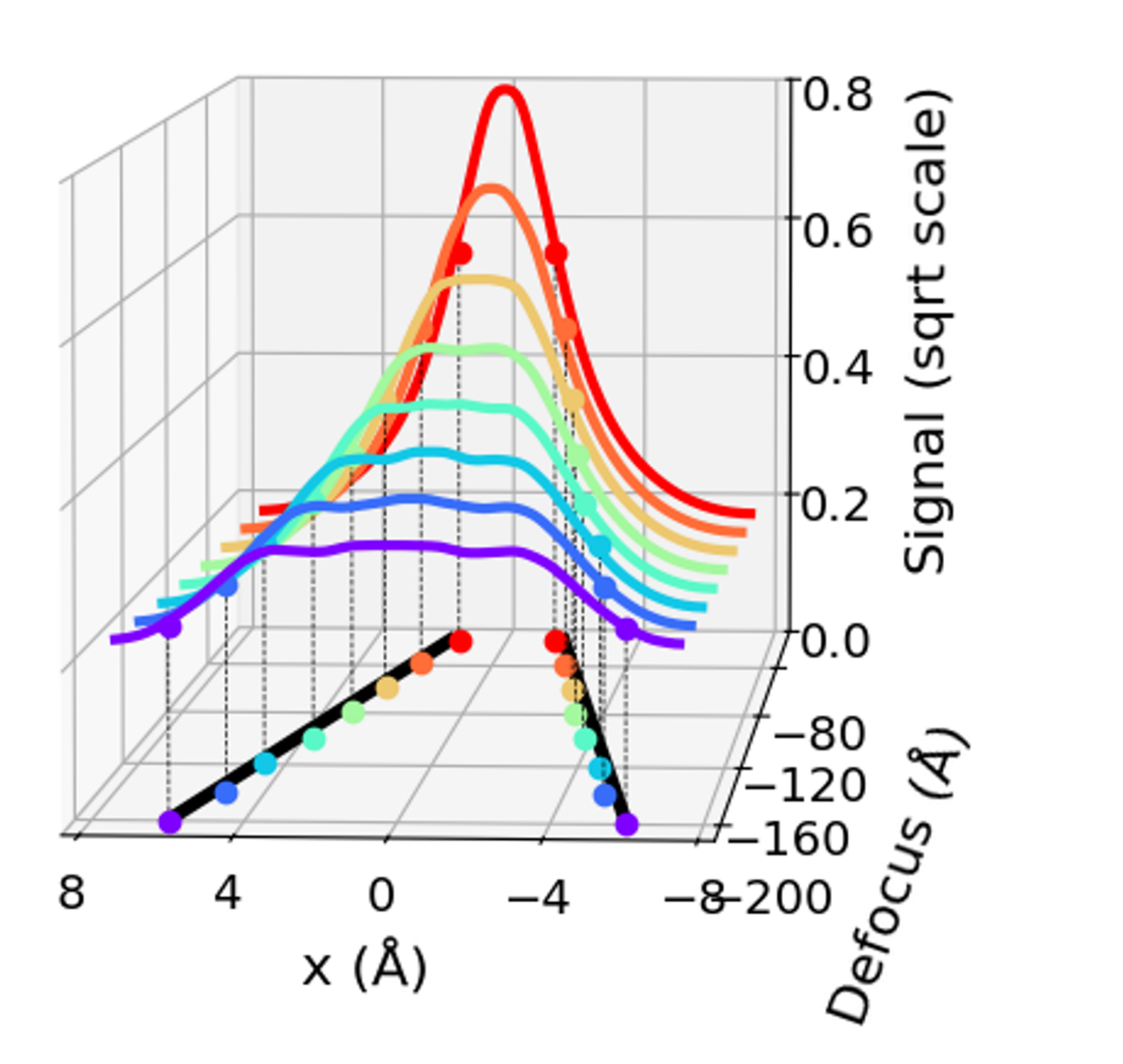}
    \caption{HWHM-predicted defocus versus simulated defocus for Si-L\textsubscript{2,3} at 80 keV. Each colored curve shows a line profile through a simulated single-atom STEM-EELS image at a given defocus (red: small underfocus, purple: large underfocus). Colored dots show the defocus predicted from the HWHM of each profile, which closely follows the true simulated defocus (black lines), supporting HWHM as a proxy for dopant depth.}
    \label{fig:dopant_depth_using_HWHM}
\end{figure}

We approximate the intensity described by the convolution model in \cref{eq:conv_model} using the super-Gaussian model defined in \cref{eq:sup_gaussian}. %Because this empirical model does not contain an explicit radius parameter, we instead used its full width at half maximum (HWHM), which we hypothesized would be consistent with the dopant depth. 
We claim that the super-Gaussian half width at half maximum (HWHM) radius $r_i$ in \cref{eq:sup_gaussian} can be used when deducing dopant depth by geometric reasoning in \cref{eq:rad}. To substantiate this, \cref{fig:dopant_depth_using_HWHM} compares the measured HWHM on Si-L$_{2,3}$ defocused STEM-EELS image line scans (shown by the coloured dots) at various defocii with the theoretically expected defocus, equal to the dopant's distance from the focal plane (shown in black). The two are in good agreement, supporting the validity of this approximation.

%% file: supplementary/Background_Amplitude_using_MLE.tex
\section{Spectrum background amplitude using maximum likelihood estimation (MLE)}\label{sec:background-amplitude-mle}

While the amplitude of the EEL spectrum edge background can be estimated using conventional least-squares fitting, MLE provides a statistically rigorous alternative, particularly in the low-count regime where the assumptions underlying least-squares fitting may no longer hold.

In the pre-edge window the expected counts at pixel $(x,y)$ are $a_{xy} b_E $ where $b_E=E^{-\gamma}$. The Poisson likelihood (by analogy to \cref{eq:ll_img} of the main manuscript but restricted to a fixed probe position) is 
\begin{align}
\mathcal{L} =
\prod_{E\in\mathrm{pre}}
\frac{
\left[a_{xy} \hat b_E\right]^{D_{xyE}} e^{-a_{xy}\hat  b_E}
}{
D_{xyE}!
} \;,
\end{align}
where $D_{xyE}$ is the spectrum counts at pixel $(x,y)$. The Poisson log-likelihood over the pre-edge window is then
\begin{equation}
    \ell(a_{xy}) = \Bigl(\sum_{E \in \mathrm{pre}} D_{xyE}\Bigr)\log a_{xy}
    - a_{xy}\sum_{E \in \mathrm{pre}} \hat b_E + \mathrm{const}.
\end{equation}
Setting $\partial \ell / \partial a_{xy} = 0$ gives the closed-form MLE

\begin{equation}
    \hat{a}_{xy} = \frac{\displaystyle\sum_{E \in \mathrm{pre}} D_{xyE}}{\displaystyle\sum_{E \in \mathrm{pre}} \hat{b}_E} = \frac{\displaystyle\sum_{E \in \mathrm{pre}} D_{xyE}}{B}
\end{equation}

where we have defined $B = \sum_{E\in\mathrm{pre}} \hat{b}_E$.  The estimator is unbiased since
\begin{align}
    \mathbb{E}[\hat a_{xy}]
= \mathbb{E}\!\left[\frac{\sum_{E\in\mathrm{pre}} D_{xyE}}{B}\right]
= \frac{1}{B}\sum_{E\in\mathrm{pre}}\mathbb{E}[D_{xyE}]
= \frac{1}{B}\sum_{E\in\mathrm{pre}} a_{xy}b_E
= a_{xy},
\end{align}
and has variance

\begin{align}
\mathrm{Var}(\hat{a}_{xy})
&= \mathrm{Var}\left(\frac{\sum_{E\in\mathrm{pre}} D_{xyE}}{B}\right) = \frac{1}{B^2}\sum_{E\in\mathrm{pre}}\mathrm{Var}(D_{xyE}) \\
&= \frac{1}{B^2}\sum_{E\in\mathrm{pre}}a_{xy}b_E = \frac{a_{xy}}{B}
\approx \frac{\hat{a}_{xy}}{B}.
\label{eq:supp_bkg_fit_var}
\end{align}

where the second line follows from Poisson counting statistics. %consistent with \cref{eq:amp_mle}. The variance is inversely proportional to the total pre-edge counts $B$, and is therefore negligible for pre-edge windows of 100\,eV or more.

% A verification of these results is provided in \cref{fig:MLE-bkg-fit-dist}. A background with power-law of the form $9\times 10^{10} E^{-2.5}$ is sampled \hl{elaborate: energy window? number of bins?} with Poisson noise and the amplitude is recovered repeatedly \hl{what does that mean? Is `repeatedly' just the recovery, or the generation of different noise realisations?} via MLE, producing the histograms. The true amplitude is subtracted so that each distribution is centred about zero. The theoretically predicted Gaussian is overlaid as a dashed line. \hl{More explanation required, and a discussion of what the figure shows.}

\begin{figure}
    \centering
    \includegraphics[width=0.75\linewidth]{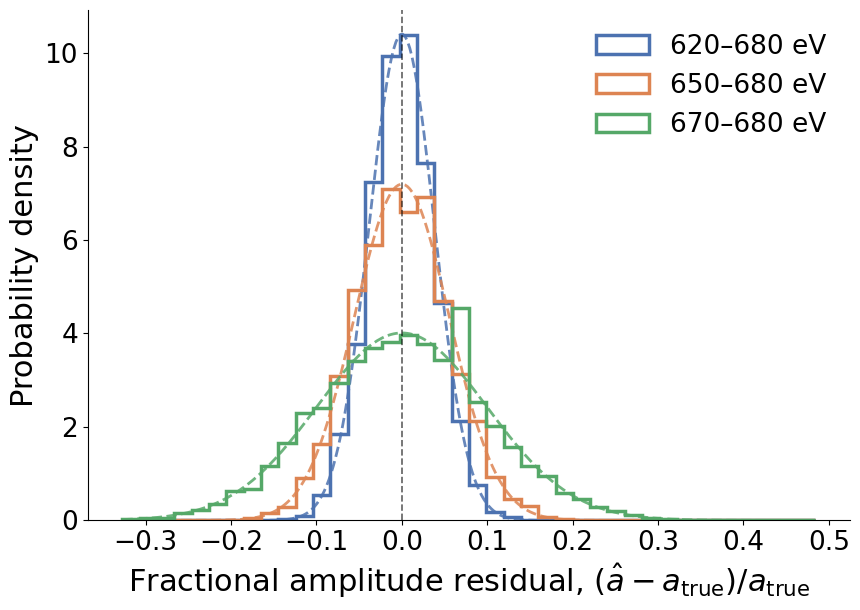}
    \caption{
    Distribution of the fractional residual $(\hat{a} - a_\mathrm{true})/a_\mathrm{true}$ having applied MLE to $10^4$ independent Poisson noise realisations of the power-law background ($a_\mathrm{true} = 1.5 \times 10^8$, $\gamma = 2.5$) for three different pre-edge window ranges. All are centred on zero, confirming the estimator is unbiased. Their variance decreases with narrower pre-edge window, consistent with reduced counts constraining the estimate. All agree well with the theoretically predicted Gaussians (dashed lines). 
    %To verify the MLE amplitude estimator, $10^4$ independent Poisson realisations of a power-law background ($a_\mathrm{true} = 8.951 \times 10^{10}$, $\gamma = 2.5463$) \hl{these number are not very round --- where did they come from?} were generated and fitted independently for three pre-edge windows of decreasing width (620--680\,eV, 650--680\,eV, 670--680\,eV). The resulting distributions of the fractional residual $(\hat{a} - a_\mathrm{true})/a_\mathrm{true}$ are all centred on zero, confirming the estimator is unbiased, and agree well with the theoretically predicted Gaussians (dashed lines). The variance increases as the fitting window narrows, consistent with fewer counts constraining the estimate.
    % $a=15 \times 10^7$, $A_{\rm true}=8.951 \times 10^{10}$, $\gamma$=2.5463
    }
    \label{fig:MLE-bkg-fit-dist}
\end{figure}

A verification of these results is provided in \cref{fig:MLE-bkg-fit-dist}. Synthetic spectra are generated from a power-law background $b(E) = a_\mathrm{true}E^{-\gamma}$ with $\gamma = 2.5$ and $a_\mathrm{true} = 1.5\times10^{8}$, the amplitude fitted to a representative pixel of the experimental dataset. The energy dispersion per-bin is held fixed at  $\Delta E=0.9$~eV while the background window range is increased. The Fe-L$_{2,3}$ edge window of 690-710 eV corresponds to $\sim 231$ background counts or $\sim$10  background counts per bin.  

Ten thousand ($10^4$) independent Poisson realisations of the same underlying model, $b(E)$, were generated, each fitted independently, and the resulting $\hat{a}$ values are summarised as a histogram in \cref{fig:MLE-bkg-fit-dist} for three different fitting windows of widths 60 eV, 30 eV and 10 eV, and all ending at 680 eV. %beginning at 620, 650, 670 eV using independent sets of realisations.
The true amplitude is subtracted so that each distribution is centred about zero, and the theoretically predicted Gaussian, \cref{eq:supp_bkg_fit_var}, is overlaid as a dashed line.

Two features are apparent. First, all three distributions are centred on zero, confirming that the maximum-likelihood amplitude is unbiased, i.e. no systematic offset is introduced by estimating the amplitude from a finite, noisy pre-edge region. Second, the width of the distribution grows as the fitting window is narrowed, since fewer channels, and hence fewer counts, constrain the estimate.

%% file: supplementary/Gaussian_Blurring_Effects_on_Inference.tex
% \section{Gaussian Blurring Effects on Inference}
\section{Blurring for visual denoising of STEM-EELS and score map images}\label{sec:gaussian-blurring}

As per the top row in \cref{fig:inference_showcase} of the main text, STEM-EELS images are noisy at low dose. However, since the disk structures of out-of-focus dopants exhibit extended features, it may be possible to enhance their clarity by local region averaging, which we do by Gaussian blurring. \Cref{fig:inference_showcase} shows an unblurred background-subtracted image and blurred $z$-score maps. That presentation is rounded out here by showing blurred background-subtracted images and unblurred $z$-score maps.

%Before interpreting the statistical significance of the detected contrast, it is important to emphasize that $z$-score maps are not expected to reveal contrast that is absent from a background-subtracted image. Rather, background subtraction can be viewed as the first step of hypothesis testing: although it produces similar visual contrast, the resulting intensity is not directly interpretable in terms of statistical significance. In contrast, z-score maps provide a statistically meaningful measure of that same contrast.

\begin{figure}
    \centering
    \includegraphics[width=0.9\linewidth]{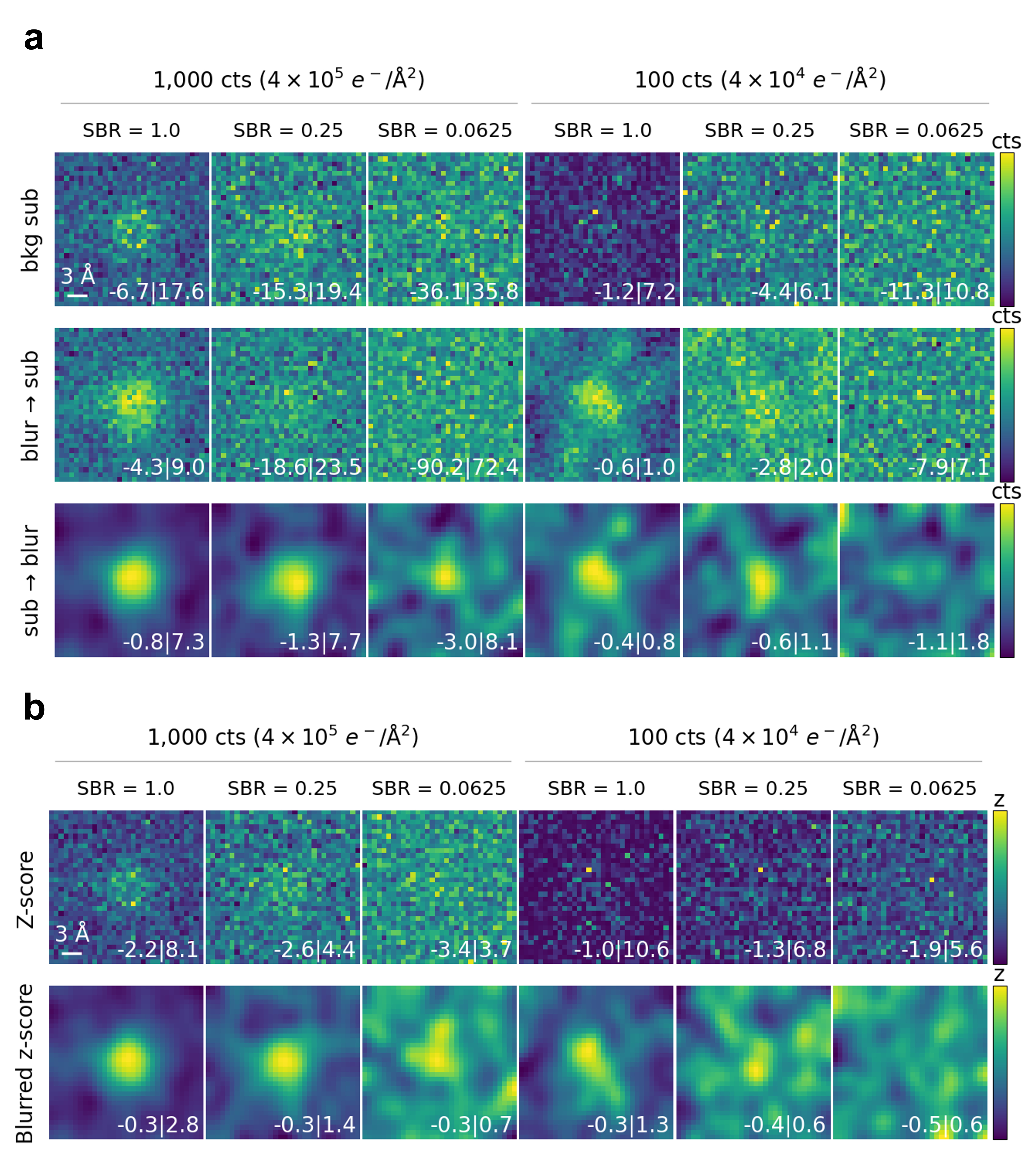}
    \caption{Comparison of background-subtraction strategies and $z$-score maps across dose ($1{,}000$ and $100$ counts per pixel) and signal-to-background ratio (SBR). {\bf a} Unblurred background-subtracted images (top) show little dopant contrast. Applying a  2${\rm\AA}$ standard deviation Gaussian blur, before (middle) or after (bottom) background subtraction improves dopant visibility. {\bf b} Unblurred $z$-score maps (top) similarly show little dopant contrast, but blurred z-score maps (bottom) using 2${\rm\AA}$ standard deviation Gaussian blur recover comparable contrast to the blurred, background-subtracted images in {\bf a}.  Numbers inset indicate the min$|$max value.}
    \label{fig:additonal_inference_tests}
\end{figure}

\Cref{fig:additonal_inference_tests}{\bf a} compares three background-subtraction strategies. The first row shows background subtraction without blurring, the second row shows the raw counts blurred before background subtraction, and the third row shows blurring applied after background subtraction. The negligible difference between blurring applied before or after background subtraction confirms that the sequence of operations does not significantly affect the result. Blurring substantially enhances the visibility of the contrast, producing images in which dopant visibility is comparable to that in the blurred $z$-score maps shown in the main manuscript (reproduced in the bottom row of \cref{fig:additonal_inference_tests}(b) for reference). However, blurred background-subtracted image intensity is not directly interpretable in terms of statistical significance. The $z$-score maps overcome that deficit, providing a statistically meaningful measure of that same contrast.

%However, unlike the $z$-score maps, the blurred background-subtracted images do not distinguish genuine signal from fluctuations arising from noise \hl{maybe not, but when integrated over energy loss do we not expect those noise fluctuations to be appreciably suppressed?}. 

Because the $z$-score analysis is carried out on a pixel-by-pixel basis, the unblurred $z$-score maps shown in top row of \cref{fig:additonal_inference_tests}{\bf b} contain little visually interpretable contrast. The key advantage of the $z$-score approach is that the contrast is statistically interpretable: each pixel value directly reflects the number of standard deviations by which the signal exceeds the background, enabling principled detection of the dopant. Putting this together, we conclude that the combination of statistical normalization and spatial blurring is required to produce maps that are both visually interpretable and statistically meaningful.

% \begin{figure}
%     \centering
%     \includegraphics[width=1\linewidth]{figures/v5/suppl_blur_bkg_sub.png}
%     \caption{Caption}
%     \label{fig:blur_bkg_subbed}
% \end{figure}

% \begin{figure}
%     \centering
%     \includegraphics[width=1\linewidth]{figures/v5/suppl_z_score_unblur_comp.png}
%     \caption{Caption}
%     \label{fig:z_score_comp}
% \end{figure}

%% file: supplementary/score_test_derivation.tex
\section{Score-test derivation}\label{sec:score-test}

At each pixel $(x,y)$ we compare the hypotheses:
\begin{align*}
    H_0 &: I(\mathbf{R})+c = 0 \quad\text{(background only)},\\
    H_1 &: I(\mathbf{R}) +c> 0 \quad\text{(edge signal present)}.
\end{align*}
where $c\geq 0$ is some constant offset.
The counts within a single voxel are given by \cref{eq:eels_model} and, as such, the Poisson likelihood for a single pixel is given by  
\begin{align}
     \mathcal{L} =
\frac{
\left[\hat a_{xy} \hat b_E + (I(\mathbf{R})+c)\,\hat t_E \right]^{D_{xyE}} e^{-[\hat a_{xy} \hat b_E+ (I(\mathbf{R})+c)\,\hat t_E]}
}{
D_{xyE}!
} \;,
\end{align}
Over the edge window the log-likelihood is given by 
\begin{equation}
    \ell(a_{xy}, I(\mathbf{R})+c) = \sum_{E \in \mathrm{edge}}
    \Bigl[D_{xyE}\,\log\!\bigl(\hat{a}_{xy}\hat{b}_E + (I(\mathbf{R})+c)\,\hat{t}_E\bigr)
    - \bigl(\hat{a}_{xy}\hat{b}_E + (I(\mathbf{R})+c)\,\hat{t}_E\bigr)\Bigr] + \mathrm{const}.
\end{equation}
Rather than maximising over $I(\mathbf{R})+c$ under $H_1$, the score test (Rao test) evaluates the
gradient of $\ell$ with respect to $I(\mathbf{R})+c$ at the null-hypothesis, $\ell(a_{xy}, I(\mathbf{R})+c=0)$. A large
positive gradient at the null indicates that increasing $I(\mathbf{R})+c$ from zero would improve the
fit, i.e.\ that the data favour $H_1$. Differentiating,
\begin{align}
    U_{xy} &= \left.\frac{\partial \ell}{\partial (I(\mathbf{R})+c)}\right|_{I(\mathbf{R})+c=0}
    \nonumber\\[4pt]
    &= \sum_{E \in \mathrm{edge}}
    \frac{D_{xyE}\,\hat{t}_E}{\hat{a}_{xy}\hat{b}_E} - \sum_{E \in \mathrm{edge}} \hat{t}_E.
    \label{eq:inter_Uxy}
\end{align}
Let $\Delta_{xyE} = D_{xyE} - \hat{a}_{xy}\hat{b}_E$ be the per channel residual,
then \cref{eq:inter_Uxy} simplifies to
\begin{equation}
    U_{xy} = \sum_{E \in \mathrm{edge}}
    \frac{\Delta_{xyE}\,\hat{t}_E}{\hat{a}_{xy}\,\hat{b}_E},
\end{equation}
which can be interpreted as a weighted template match where the background-subtracted spectrum at pixel
$(x,y)$ is projected onto the edge shape $\hat{t}_E$, with each channel weighted inversely by the
local background level $\hat{a}_{xy}\hat{b}_E$.

Under $H_0$, $D_{xyE} \sim \mathrm{Poisson}(\hat{a}_{xy}\hat{b}_E)$, so
$\mathrm{Var}(D_{xyE}) = \hat{a}_{xy}\hat{b}_E$ and channels are independent. Propagating through
the expression for $U_{xy}$,
\begin{align}
    \mathrm{Var}(U_{xy})
    &= \sum_{E \in \mathrm{edge}}
    \frac{\hat{t}_E^2}{\hat{a}_{xy}^2\hat{b}_E^2}\,\mathrm{Var}(D_{xyE})
    \nonumber\\[4pt]
    &= \sum_{E \in \mathrm{edge}}
    \frac{\hat{t}_E^2}{\hat{a}_{xy}^2\hat{b}_E^2}\,\hat{a}_{xy}\hat{b}_E
    = \sum_{E \in \mathrm{edge}}
    \frac{\hat{t}_E^2}{\hat{a}_{xy}\,\hat{b}_E}.
\end{align}

Normalizing the score by its standard deviation under $H_0$ yields the standardised statistic
\begin{equation}
    z_{xy} = \frac{U_{xy}}{\sqrt{\mathrm{Var}(U_{xy})}}
    = \frac{\displaystyle\sum_{E \in \mathrm{edge}}
            (D_{xyE} - \hat{a}_{xy}\hat{b}_E)\,\dfrac{\hat{t}_E}{\hat{b}_E}}
           {\sqrt{\hat{a}_{xy}
            \displaystyle\sum_{E \in \mathrm{edge}}
            \dfrac{\hat{t}_E^2}{\hat{b}_E}}},
\end{equation}
which result is given in \cref{eq:zmap} of the main manuscript.

%% file: supplementary/inference_strategy.tex
\section{Inference strategy with multiple dopants}\label{sec:inference-strategy}

When multiple dopants are present, parameterizing disk radii directly introduces an exchange symmetry where permutations of the disk labels correspond to equivalent solutions, resulting in a multi-modal posterior that is costly to sample. We remove this degeneracy by parameterizing radii cumulatively, i.e. with $\delta_i$ such that $\delta_1=r_1, \delta_2=r_2-r_1,...$. For example, radii (5,10,20)Å are represented as (5,5,10)Å, i.e. the first radius followed by successive increments. A positivity constraint is then imposed, ensuring a unique ordering of the disks and eliminating the exchange symmetry.

%% file: supplementary/Examining_FOV.tex
\section{Further supporting experimental data}\label{sec:examining-fov}

% \begin{figure}
%     \centering
%     \includegraphics[width=0.75\linewidth]{figures/v4/experiment/pacbed.png}
%     \caption{(a) Experimental (b) simulated 7.8nm thick}
%     \label{fig:placeholder}
% \end{figure}

\begin{figure}
    \centering
    \includegraphics[width=0.7\linewidth]{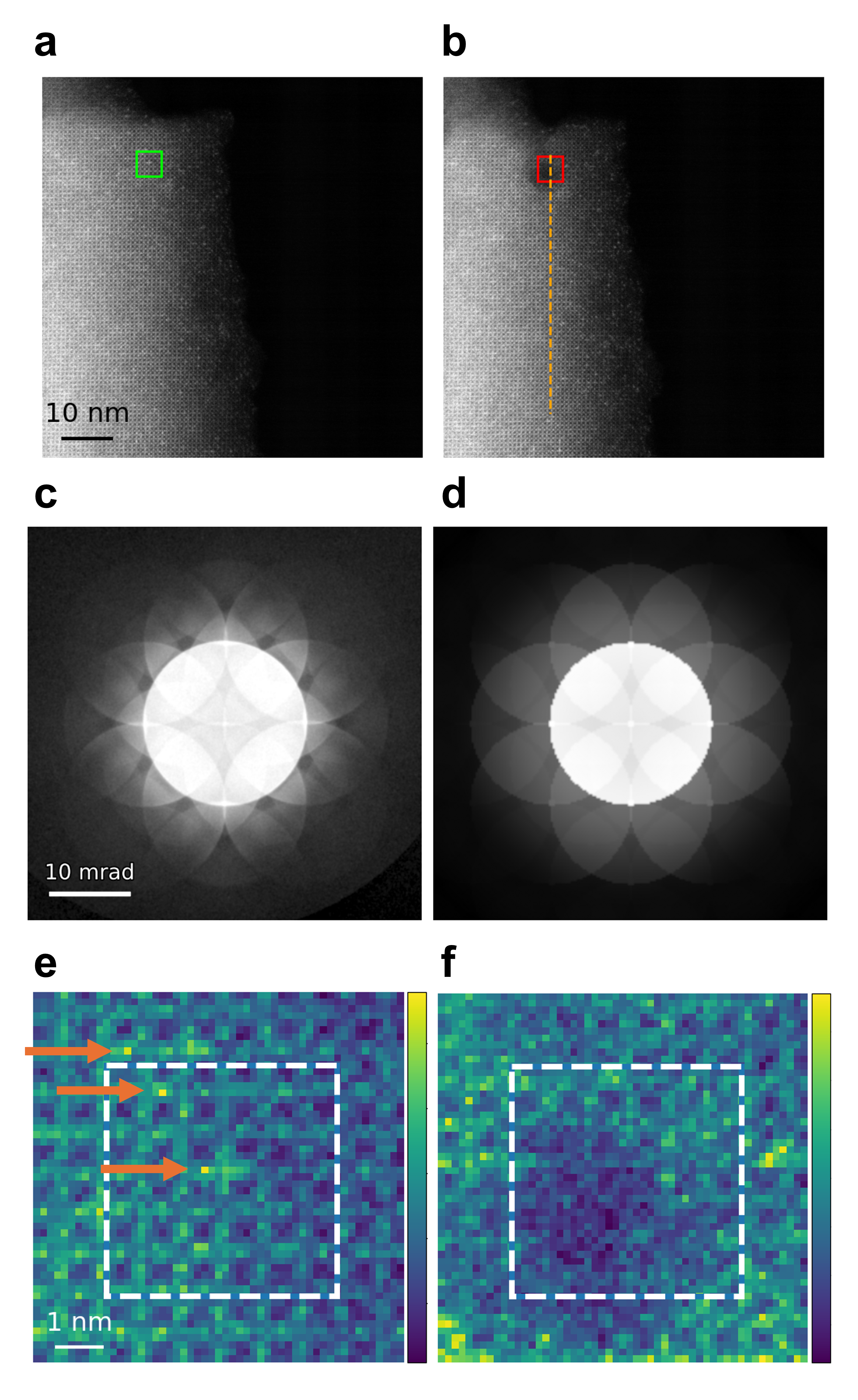}
    \caption{HAADF survey images acquired {\bf a} before and {\bf b} after the defocused STEM-EELS acquisition, with the acquisition region indicated by the coloured boxes. {\bf c} Experimental PACBED pattern acquired over an extended region of approximately uniform thickness in the vicinity of the orange reference line in {\bf b}. {\bf d} Simulated PACBED pattern for a sample thickness of 7.8\,nm. Magnified views of the acquisition region {\bf e} before and {\bf f} after acquisition, with the STEM-EELS acquisition region indicated by the dashed white box.}% Two dopants are visible in (e), while a crater is apparent in (f) at the location of the central dopant, suggesting it was ejected during acquisition.}
    \label{fig:fov_exam}
\end{figure}

High angle annular dark field (HAADF) surveys of the selected spinel shard before and after the defocused STEM-EELS acquisition are shown in \cref{fig:fov_exam}{\bf a} and {\bf b}, respectively. The change in contrast in the EELS acquisition region, marked by the squares, shows that appreciable beam damage occurred during acquisition.

Prior to STEM-EELS acquisition, a PACBED pattern, shown in  \cref{fig:fov_exam}{\bf c}, was acquired over the vicinity of  the orange reference line shown in \cref{fig:fov_exam}{\bf b}, an extended region of approximately uniform thickness. The appearance and features in that experimental PACBED pattern are in good agreement with the simulated PACBED pattern in \cref{fig:fov_exam}{\bf d} assuming a sample thickness of 7.8 nm, which we therefore take to be a good estimate of the sample thickness in the EELS acquisition region.

\Cref{fig:fov_exam}{\bf e} and {\bf f} show magnified sections of \cref{fig:fov_exam}{\bf a} and {\bf b} around the STEM-EELS acquisition region, denoted by the white dashed rectangle. Before acquisition (\cref{fig:fov_exam}{\bf e}), two dopants in the field of view are visible as points of high HAADF contrast within the EELS scan acquisition region: one near the upper left and another closer to the centre. A third dopant is present just outside the upper left corner. (All three are indicated with arrows in \cref{fig:fov_exam}{\bf e}.) None are evident after acquisition (\cref{fig:fov_exam}{\bf f}). The central dopant is absent from the $z$-score map in \cref{fig:experiment_analysis}, suggesting that it became mobile during acquisition, and may be contributing to the diffuse contrast in the $z$-score map that indicates presence of Fe not conforming to the disk-like contrast expected of a fixed dopant. %Consistent with this interpretation, the post-acquisition image shows a crater at approximately the same location, indicating that the dopant was likely ejected. 
However, the upper left dopant is evident in the $z$-score map in \cref{fig:experiment_analysis}, and, as discussed further in \cref{sec:verifying-inference}, its appearance agrees well with the disk-like contrast of a fixed dopant. We therefore interpret that its displacement occurred late in the acquisition, with little adverse impact on its detected signal.

\begin{figure}
    \centering
    \includegraphics[width=1\linewidth]{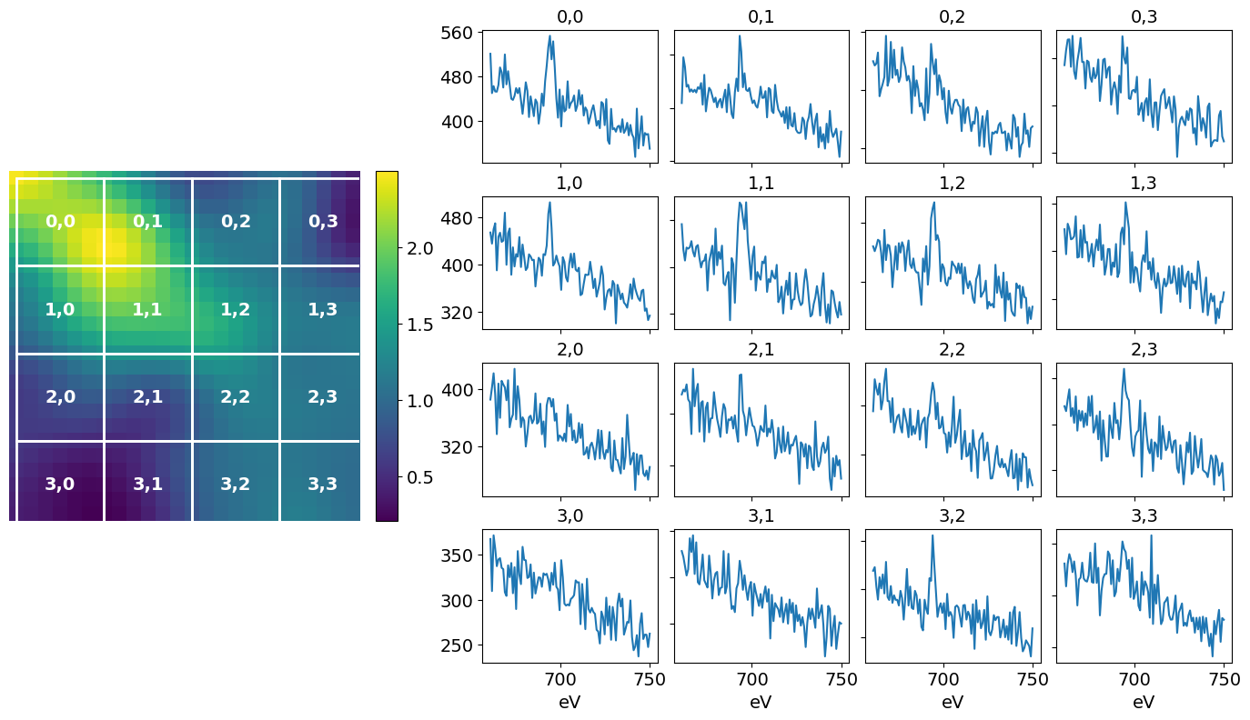}
    \caption{Spatially resolved EEL spectra (right) extracted from a $4\times4$ spatial binning of the acquisition field of view, with bin positions indicated on the $z$-score map with 5${\rm\AA}$ standard deviation Gaussian blur (left). Each spectrum shows the cumulated counts within that binned region. Spectra from tiles near the dopant in the top left show a visible edge onset above the power-law background, consistent with the elevated $z$-scores at those positions, while spectra from bottom left show a smoothly decaying background with no discernible edge, consistent with the absence of dopant signal there. The spectra corresponding to tiles with second index $\geq 2$ show the dopant signal is clearly present, consistent with the $z$-score map.}
    \label{fig:suppl_spectrum_regions}
\end{figure}

The contrast in the $z$-score map is corroborated by examining the spectra from bulk regions of the sample in \cref{fig:suppl_spectrum_regions}.
To the left is the $24 \times 24$ pixel $z$-score map which has been binned into $6\times 6$ regions, with the integrated spectrum from each bin shown in a corresponding tile on the right to further examine the correspondence between contrast in the $z$-score map and the underlying dopant signal.
The presence of peaks is indeed consistent with the $z$-score map.  (Discrepancy in edge onset is suspected to be due to the non-linear dispersion across the energy-loss spectrum from the large dispersion of 0.9eV per channel used.) The $z$-score map also indicates a dopant just beyond the upper-left corner of the field of view, which is consistent with in the pre-acquisition HAADF image in \cref{fig:fov_exam}{\bf e}.

% \begin{figure}
%     \centering
%     \includegraphics[width=0.5\linewidth]{figures/v4/experiment/survey.png}
%     \caption{Caption}
%     \label{fig:haadf_survey}
% \end{figure}

%% file: supplementary/Verifying_Inference_Results.tex
\section{Corroborating $a$ $posteriori$ model parameter fits}\label{sec:verifying-inference}
% \label{sec:corr_a_post}
% keep italics?

\begin{figure}
    \centering
    \includegraphics[width=1\linewidth]{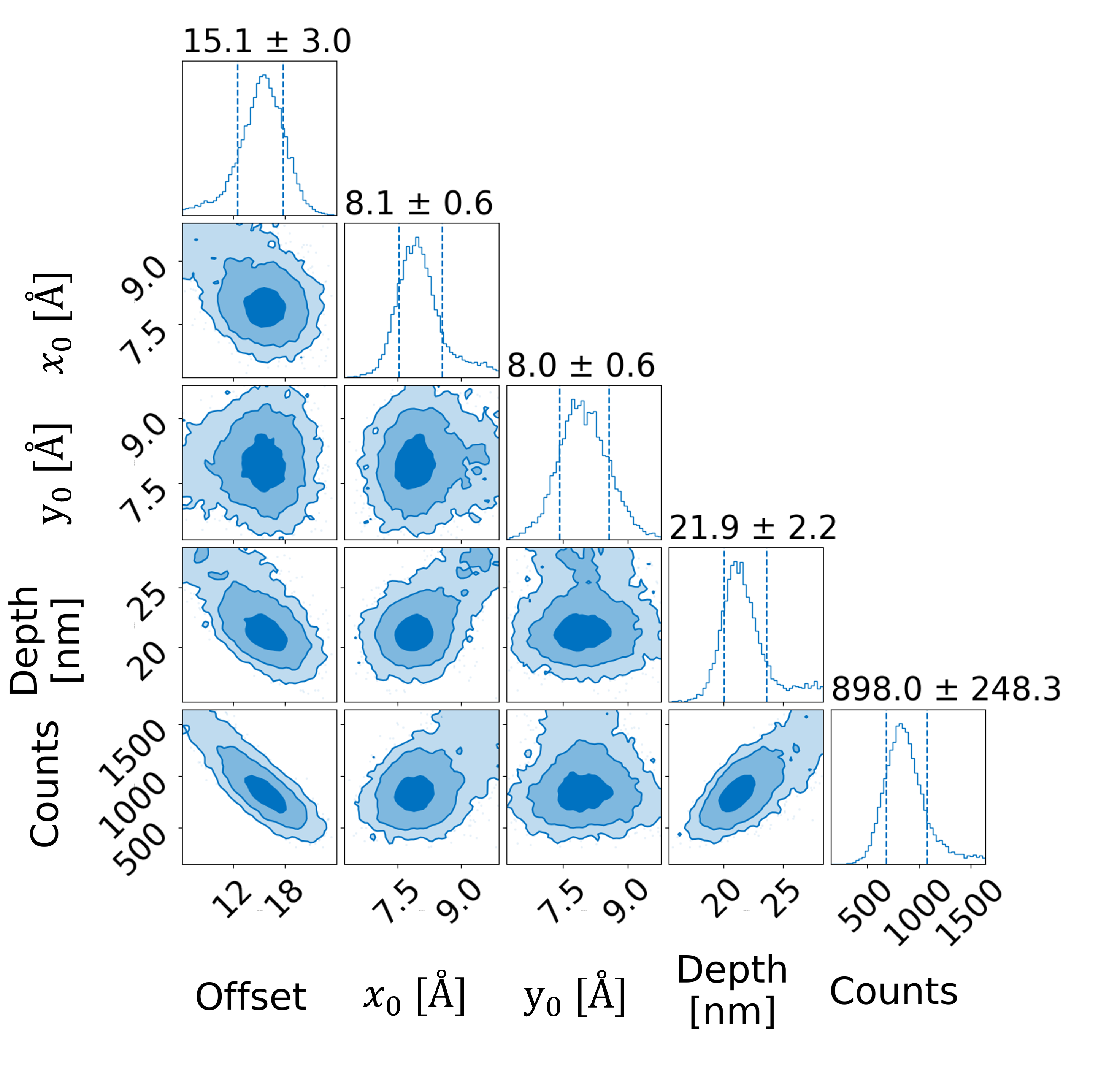}
    \caption{Full posterior distribution of the Bayesian inference parameters: 
background offset, dopant position ($x_0$, $y_0$), depth, and total signal 
counts. Diagonal panels show the marginal distributions with mean and standard 
deviation annotated overtop. Off-diagonal panels show pairwise joint distributions.}
    \label{fig:full_posterior}
\end{figure}

A subset of the model parameter posteriors is presented in the main manuscript. The full posterior distribution is shown in \cref{fig:full_posterior}. As discussed, the inferred disk centre and radius are consistent with both the $z$-score map and the experimental geometry. In this section we present additional consistency tests supporting the accuracy of model parameters deduced from the posterior, which further bolsters our confidence in the deduced dopant depth.%\hl{As a further consistency test, to ensure the posterior obtained is wholly sensible is to obtain independent, approximate estimates of the background offset and the signal counts. If the posterior is found to be in predicting all model parameters, then this increases our confidence in the obtained dopant depth.}

The background offset, {i.e.} $c$ in \cref{eq:eels_model}, can be estimated from the bottom-right region of the field of view in \cref{fig:experiment_analysis}{\bf a}, denoted by $\Omega_{\mathrm{BR}}$, which contains a nearly uniform signal with no visible dopants. The offset is approximately

\begin{align}
\mathrm{offset} \approx \frac{1}{|\Omega_{\mathrm{BR}}|} \sum_{(x,y)\in\Omega_{\mathrm{BR}},, E\in\mathrm{edge}} \left(D_{xyE} - \hat{a}_{xy} b_E\right) = 12,
\end{align}
where $|\Omega_{\mathrm{BR}}|$ denotes the number of pixels in region $\Omega_{\mathrm{BR}}$. The averaged residual is calculated by subtracting the fitted background from the measured data over the selected energy window. This estimate is in reasonable agreement with the posterior prediction of $15.1\pm3.0$.

Similarly, the total signal within the region used for the Bayesian inference, outlined in blue and white in \cref{fig:experiment_analysis}{\bf b} and here denoted by $\Omega_{\mathrm{TL}}$, can be estimated as

\begin{align}
\mathrm{signal} \approx \sum_{(x,y)\in\Omega_{\mathrm{TL}}, E\in\mathrm{edge}} \left(D_{xyE} - \hat{a}_{xy} b_E\right) = 2499.
 \label{eq:signal_est1}
\end{align}

% The Bayesian inference estimates that 877 of these electrons originate from the disk itself. Using the independently estimated background offset of approximately 15 electrons per pixel gives

% \begin{align}
% 15 \times |\Omega_{\mathrm{TL}}| + 877 = 2241,
% \end{align}

% which is in reasonable agreement with the direct estimate of 2499 electrons.

The Bayesian inference estimates that 898 of these electrons originate from the disk itself. Using the Bayesian-estimated background offset of approximately 15 electrons per pixel gives

\begin{align}
(15.1\pm3.0) \times |\Omega_{\mathrm{TL}}| + (898.0\pm 248) = 2248\pm367,
 \label{eq:signal_est2}
\end{align}
where $\Omega_{\rm TL}=9\times 10$, which agrees (within uncertainties) with the direct estimate of 2499 electrons of \cref{eq:signal_est1}.

\begin{figure}
    \centering
    \includegraphics[width=1\linewidth]{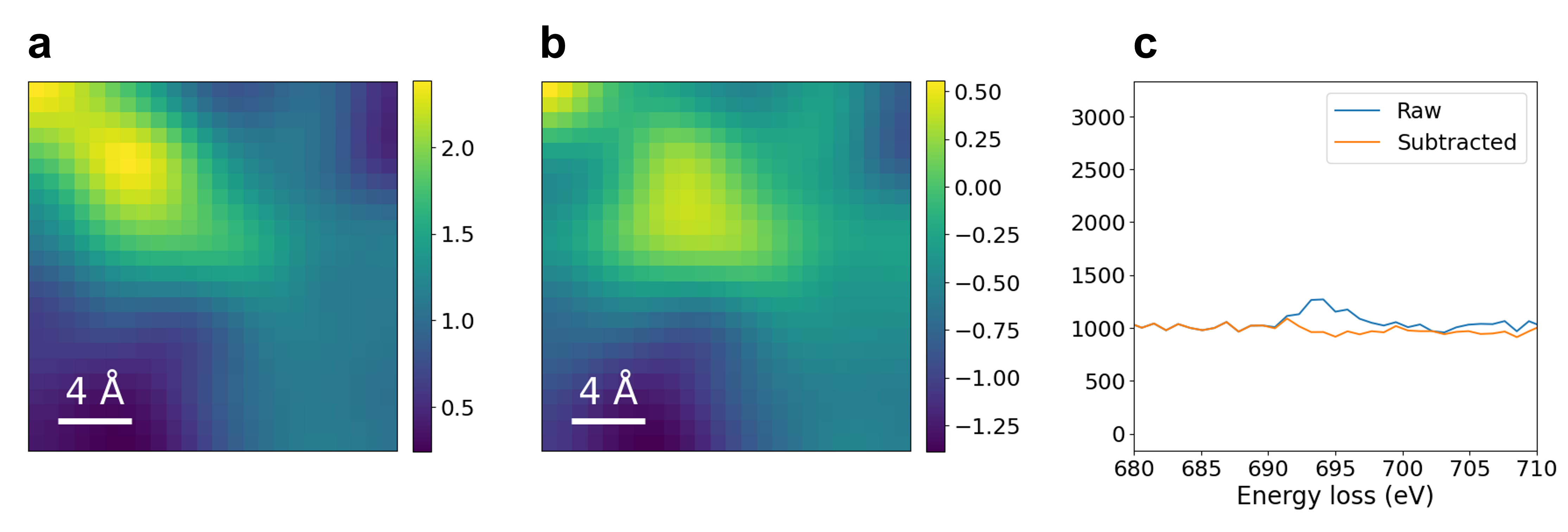}
    \caption{Validation of the inferred signal via subtraction of the maximum $a$~$posteriori$ model prediction. {\bf a} Original blurred $z$-score map. {\bf b} Residual z-score map after subtracting the inferred signal which shows a noticeable reduction in contrast. As expected, some contrast remains since Poisson noise is not additive. {\bf c} Spatially integrated spectra before (blue) and after (orange) subtraction. The edge signal is almost entirely removed in the summed spectrum, confirming that the Bayesian inference has accurately recovered the underlying edge signal.}
    \label{fig:subtracted_estimated_signal}
\end{figure}

% Additionally, we can further validate the inferred disk parameters by subtracting the intensity predicted by the model at the maximum $a$ $posteriori$ parameter estimates, 
Another check of the $a$ $posteriori$ model estimates is to subtract the per pixel predicted intensity from the dopant. \Cref{fig:subtracted_estimated_signal}{\bf a} shows the $z$-score map from the original data, and \cref{fig:subtracted_estimated_signal}{\bf b} shows the $z$-score map obtained after the dopant signal counts predicted by Bayesian estimation are subtracted from the original data. Comparing the $z$-score ranges of the two figures, 
the latter exhibits a noticeably weaker signal, although residual contrast remains. This is expected, as subtracting the mean inferred signal from individual pixels is only an approximation: the Bayesian model predicts the expected intensity, whereas each pixel is subject to Poisson counting noise. Consequently, the subtraction is not expected to perfectly remove the signal on a pixel-by-pixel basis.

In contrast, the spatially integrated spectrum contains substantially more counts, making the mean prediction a much better approximation to the observed signal. The aggregated spectra are shown in \cref{fig:subtracted_estimated_signal}{\bf c} where the dopant signal is almost completely eliminated in the summed spectrum, providing additional evidence that the Bayesian inference has accurately recovered the underlying edge signal.

\begin{figure}
    \centering
    \includegraphics[width=0.7\linewidth]{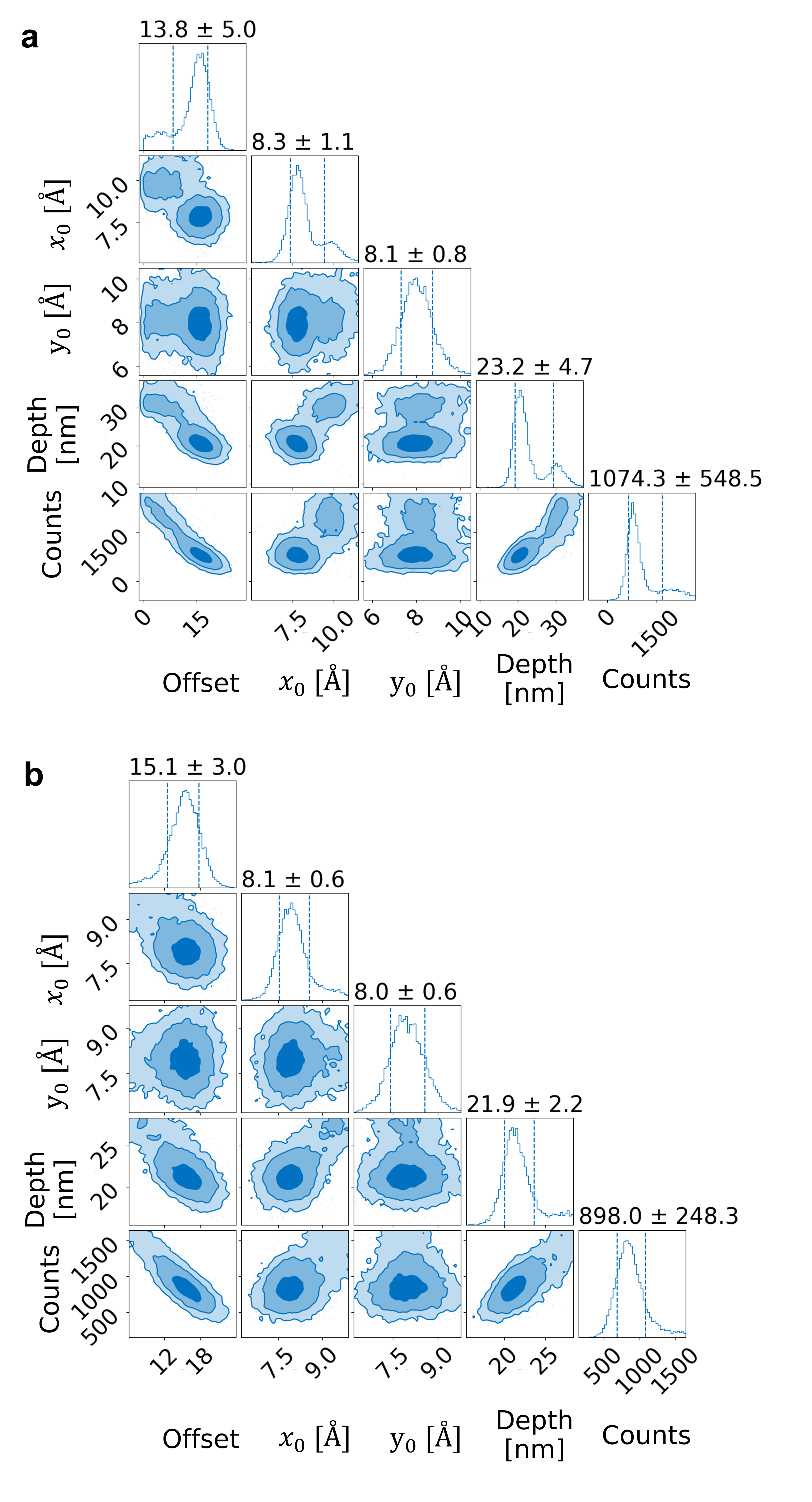}
    \caption{Comparison of posterior distributions from Bayesian inference performed on {\bf a} the spectrally integrated 2D edge image and {\bf b} the full 3D spectral data cube. Estimates of all model parameters are consistent across both the 2D and 3D models. However, inference on the 3D data cube produces consistently tighter estimates due to the additional spectral information.}
    \label{fig:2d_3d_bayesian_inference_results}
\end{figure}

As an additional validation, we repeated the Bayesian inference using the spatially integrated edge image rather than the full three-dimensional spectral data cube. In this approach, the edge intensity was first integrated over the chosen energy window to produce a two-dimensional image, and inference was performed using the corresponding image formation model. The resulting posterior distributions are compared with those obtained from the full spectral data cube in \cref{fig:2d_3d_bayesian_inference_results}. The inferred disk centre, radius, and total signal are consistent between the two approaches, indicating that the inferred parameters are robust to the choice of data representation. As expected, however, inference using the full three-dimensional data cube yields systematically tighter posterior distributions. By retaining the complete spectral information, the three-dimensional model is able to better distinguish edge signal from the background, whereas integrating over the energy window discards spectral information and consequently reduces the amount of information available for inference.

% \begin{figure}
%     \centering
%     \includegraphics[width=1\linewidth]{figures/v5/experiment_2d_vs_3d_inference.png}
%     \caption{Caption}
%     \label{fig:2d_3d_bayesian_inference_results}
% \end{figure}

